\documentclass[aps,prx,preprint,superscriptaddress]{revtex4-2}
\RequirePackage{lineno}
\usepackage[symbol]{footmisc}
\usepackage{graphicx}
\usepackage{dcolumn}
\usepackage{amsmath}
\usepackage{amssymb}
\usepackage{bm}
\usepackage{pifont}
\usepackage{epsfig}
\usepackage{psfrag}
\usepackage{epstopdf}
\usepackage{wasysym}
\usepackage[usenames]{color}
\usepackage{setspace}
\usepackage{hyperref}
\usepackage{wrapfig}
\usepackage[normalem]{ulem}
\usepackage{xcolor}
\usepackage{lineno}
\graphicspath{ {./figs/}}
\usepackage{chemformula}

\hypersetup{colorlinks = true, citecolor=blue, linkcolor=red, urlcolor=blue}

\begin{document}

\title{
Revealing hidden medium-range order in silicate glass-formers using many-body correlation functions
}

\author{Zhen Zhang}
\email[Corresponding author: ]{zhen.zhang@cdut.edu.cn}
\affiliation{Department of Physics, College of Mathematics and Physics, Chengdu University of Technology, Chengdu 610059, China}

\author{Walter Kob}
\email[Corresponding author: ]{walter.kob@umontpellier.fr}
\affiliation{Department of Physics,
University of Montpellier and CNRS, F-34095 Montpellier, France}
\date{\today}

\begin{abstract}  
The medium range order (MRO) in amorphous systems has been linked to complex features such as the dynamic heterogeneity of supercooled liquids or the plastic deformation of
glasses. However, the nature of the MRO in these materials has remained elusive, primarily due to the lack of methods capable of characterizing this order. 
Here, we leverage standard two-body structural correlators and advanced many-body correlation functions to probe numerically the MRO in prototypical network glassformers, i.e., silica and sodium silicates, systems that are of great importance in natural as well as industrial settings.
With increasing Na concentration, one finds that the local environment of Na becomes more structured and the spatial distribution of Na on intermediate length scales changes from blob-like to channel-like, indicating a growing inhomogeneity in the spatial Na arrangement. In parallel, we find that the Si-O network becomes increasingly depolymerized, resulting in a ring size distribution that broadens. The radius of gyration of the rings is well described by a power-law with an exponent around 0.75, indicating that the rings are progressively more crumbled with increasing size. Using a recently proposed four-point correlation function, we reveal that the relative orientation of the tetrahedra shows a surprising transition at a distance around 4~\AA, a structural modification that is not seen in standard two-point correlation functions. The order induced by this transition propagates to larger distances, thus affecting the structure on intermediate length scales. Furthermore, we find that, for Na-rich samples the length scale characterizing the MRO is non-monotonic as a function of temperature, caused by the competition between energetic and entropic terms which makes that the sample forms complex mesoscpic domains. Finally, we demonstrate that the structural correlation lengths as obtained from the correlation functions that quantify the MRO are correlated with macroscopic observables such as the kinetic fragility of the liquids and the elastic properties of the glasses. These findings allow to reach a deeper understanding of the nature of the MRO in network glassformers, insight that is crucial for establishing quantitative relations between their MRO and macroscopic properties.

\end{abstract}

\maketitle

\section{Introduction}
Understanding the structure of disordered systems such as liquids and glasses, and establishing its relation with the macroscopic properties of the material is of fundamental as well as technological importance~\cite{binder_kob_2011,hansen_theory_2013,varshneya2013fundamentals}. Depending on the length scale considered, structural information can be categorized into short-range order (SRO), medium (or intermediate)-range order (MRO), and long-range order (which in glassy systems is often considered to be negligible)~\cite{elliott_medium-range_1991}. Usually, SRO deals with structures within the first two nearest neighbor shells and is directly related to the nature of the particle interactions (ionic, covalent, etc.). Many indicators that characterize the SRO have been proposed (e.g., locally favored structures in hard-sphere-like systems and tetrahedral order in open-network systems) and correlations between these quantities and the macroscopic properties of the systems, such as the relaxation dynamics of the liquid, have been documented~\cite{coslovich_understanding_2007,geske2016fragile,royall_role_2015,tanaka_revealing_2019}. 
However, despite these identified correlations, there is increasing evidence that the SRO alone is insufficient for rationalizing many thermo-physical properties of glassy systems.  This indicates that structural information on larger length scales, notably the MRO, needs to be taken into account to understand properties such as dynamic heterogeneity in liquids~\cite{kawasaki2007correlation,watanabe2008direct},  kinetic 
fragility~\cite{mauro_structural_2014, yu2022silica,sidebottom2019fragility,ryu2020origin}, glass formation ability~\cite{fan2020unveiling,wang2019hints}, structural heterogeneity in glasses ~\cite{zhang2013calorimetric,reibstein2011structural,kirchner2022beyond}, or elastic properties of glasses~\cite{rouxel_elastic_2007,greaves_poissons_2011,liu2014abnormal}. 
However, the quality of the correlation identified so far depends strongly on the system and property investigated and a real understanding of the nature of the MRO is therefore still lacking.

In view of the importance of the MRO, many methods and models have been developed to probe and characterize its structure. Notable examples are the static structure factor as measured in X-ray or neutron-scattering experiments (with a particular focus on the first main diffraction peak), e.g., Refs.~\cite{du2006compositional,salmon2015networks,shi2020temperature}, nuclear magnetic resonance that permits to probe the nearest neighbor connectivity of the basic structural building blocks such as [SiO$_4$] tetrahedra in silicate glasses~\cite{stebbins1992structure,angeli_insight_2011}, 
or more recently atomic electron tomography which has allowed to determine the three-dimensional (3D) atomic positions of a glass-forming alloy~\cite{yang2021determining}, and the use of persistent homology as a topological approach to analyze the MRO in atomistic simulations of amorphous solids~\cite{hiraoka2016hierarchical,sorensen2020mro}. Although these methods have proven to be useful to gain insight into the MRO structure, none of them allows to reach a deep understanding of the atomic arrangement on the scale beyond the 2-3 nearest neighbors. 

We note that the widely used two-point correlation functions, such as the structure factor, project the whole 3D structure onto one dimension, which inevitably entails a huge loss of structural information. Recovering subsequently this information is difficult since even on the local scale the particle arrangement can be surprisingly varied without a particular signature in the two-point correlation function~\cite{jonsson1988icosahedral,coslovich_understanding_2007,cheng_atomic-level_2011,royall_role_2015,
royall_locally_2017,yuan2021packing}. We also recall that a recent numerical study has demonstrated that including higher-order static and dynamic correlations in the mode-coupling theory qualitatively changes the predicted glass-transition in both fragile and strong glass formers~\cite{sciortino2001debye,luo2022many}, emphasizing the relevance of many-body correlations for dynamic properties~\cite{tanaka2020role,pihlajamaa2023emergent}. 

In view of the aforementioned shortcomings of the two-point correlation functions, it is of interest to consider higher-order correlation functions that allow to reveal directly the 3D structure on length scales beyond the SRO. One possibility to measure four-point correlation functions at intermediate and large distances has recently been proposed in Ref.~\cite{zhang2020pnas}. This approach has been applied to experimental binary granular materials as well as colloid systems~\cite{yuan2021packing,singh2023intermediate}, and has allowed to discover in these glass-formers the existence of a non-trivial MRO structure and to establish a connection between MRO and packing efficiency as well as dynamic heterogeneity. A recent numerical study using this approach has also revealed that in granular systems the presence of inter-particle friction strongly affects the intermediate-range structure~\cite{tang2023friction}.

Alkali silicates are an important subclass of oxide glasses since the presence of the alkali atoms allows to tune the properties of the glass in a way that is advantageous for applications~\cite{varshneya2013fundamentals}. For the sodium silicate systems considered in the present work, it has been speculated already long time ago that the Na atoms (also known as network modifiers) diffuse along preferential pathways, resulting in a non-trivial MRO composed of percolating zones with enhanced Na concentration, thus motivating the so-called modified random network model~\cite{greaves1985exafs}.
Although this view is compatible with various  experimental and simulation studies~\cite{ingram1989ionic,horbach2002channels,meyer_channel_2004,le2017percolation,jund2001channel,horbach_structural_2001}, other experimental and simulation investigations have given evidence for a rather homogeneous distribution of the alkali atoms in the Si-O matrix~\cite{du_medium_2004,lee2022probing}.
Further insight into the spatial extend of inhomogeneity and the nature of alkali dispersion in the structure is therefore required. 

The objective of the present work is thus to provide a comprehensive picture of the structure of silicate glass-formers containing modifiers, notably on length scales beyond the first nearest neighbor shell. In particular we will address the question how this MRO evolves with composition and temperature. To this end, we numerically probe the structure of silica and sodium silicate systems using both the standard two-body structural measures as well as the recently proposed four-point correlation functions. Our results demonstrate the presence of a structural transition and a surprisingly complex change of the structural correlation length as a function of Na concentration and temperature. Furthermore, we show that the MRO is strongly correlated with macroscopic observables concerning the kinetic fragility of the supercooled liquids and the elastic properties of the glasses, thus demonstrating the link between the MRO and the macroscopic properties of the glass-former.

\section{Methods}
\label{sec:methods}

\subsection{Simulation details}
The compositions we have investigated are pure SiO$_2$ and the binary mixture Na$_2$O-$x$SiO$_2$ (NS$x$) with $x=2, 3, 5, 10$, i.e., the mole concentration $f$ of Na$_2$O was $0 \leq f \leq 33.3$\%.
Classical molecular dynamics (MD) simulations were performed using the two-body effective potential proposed by Sundararaman \textit{et al.}~\cite{sundararaman_new_2018,sundararaman_new_2019}
which has been shown to give a reliable description of the structural and mechanical properties of sodium silicate glasses~\cite{zhang2020potential,
zhang2022stiffness,zhang2022fracture}. 
This potential consists of a short-range Buckingham term and a long-range Coulomb term, thus its functional form is given by

\begin{equation} 
V(r_{\alpha\beta}) =  \frac{q_\alpha q_\beta e^2}{4\pi \epsilon_0 r_{\alpha\beta}} +
A_{\alpha\beta} \exp(-r_{\alpha\beta}/B_{\alpha\beta}) - \frac{C_{\alpha\beta}}{r_{\alpha\beta}^6} \quad , 
\label{eq:potential}
\end{equation}

\noindent
where $r_{\alpha\beta}$ is the distance between two atoms of species $\alpha$ and $\beta$. The values of the effective charges $q_\alpha$ and potential parameters $A_{\alpha\beta}$, $B_{\alpha\beta}$, $C_{\alpha\beta}$ are given in Ref.~\cite{sundararaman_new_2019}.
The short-range potential was cut off at a distance of 8~\AA\ and the Coulombic interactions  were evaluated using the particle-particle particle mesh (PPPM) algorithm~\cite{pollock1996pppm} with an accuracy of $5\times 10^{-5}$. (Note that we have not used any cutoff for the Coulomb part, usually made at 10~\AA, in order to avoid truncation effects that could influence the MRO.)

Our samples contained typically 120,000 atoms and were in a cubic box with, at room temperature, a size around 120~\AA, 
 corresponding to the experimental glass density~\cite{bansal_handbook_1986}. Periodic boundary conditions were applied in all directions and the simulations were carried out at pressure zero. 
The samples were first melted and maintained at 3000 K for 800 ps, a time span that is sufficiently long to fully equilibrate the liquids. (At the end of these equilibration runs the mean squared displacement of Si, i.e., the slowest atomic species, was larger than 100 \AA$^2$ for all samples.) Subsequently, the melts were cooled down to 0 K with a constant cooling rate of 5 K/ps. We note that due to the high computational cost involved in the simulations of the samples of such large size with the long-range Coulombic interactions fully accounted for, we used a cooling rate that is about one-to-two orders of magnitude higher than that were used in previous simulation studies using similar potentials (for example, Refs.~\cite{li_cooling_2017,zhang2020potential}). However, as we will see later, the applied cooling rate still allowed us to reproduce faithfully the structure and other physical properties of the investigated systems and thus it can be considered to be a reasonable choice. 

To calculate the elastic properties of the glasses, we first annealed the as-quenched glass samples at 300~K for 80~ps. Subsequently, they were unaxially elongated or compressed in one
of the three axial directions with a constant strain rate of 0.5/ns, which according to our previous studies, is sufficiently small to allow convergence of the measured elastic moduli~\cite{zhang2020potential,zhang2022fracture,zhang2022stiffness}.
The box dimensions in the transverse directions were allowed to move freely to maintain zero pressure. The elastic moduli of the glasses were obtained from the stress-strain curve at strains smaller than 1\%.

All the simulations were carried out using the Large-scale Atomic/Molecular Massively Parallel Simulator software (LAMMPS)~\cite{plimpton_fast_1995} in the isothermal–isobaric ensemble at zero pressure.
Temperature and pressure were controlled using a Nos\'e-Hoover thermostat and barostat~\cite{nose_unified_1984,hoover_canonical_1985,hoover_constant-pressure_1986}. 
The time step was 1.6~fs and the temperature and pressure damping parameters were chosen to be 0.16~ps and 1.6~ps, respectively. The results presented in the following represent the average of eight independent melt-quench samples for each composition, and the error denotes the standard error of the mean, unless stated otherwise.

\subsection{Correlation functions}
For disordered systems such as liquids and glasses, the standard quantities to characterize the structure are the partial radial distribution functions $g_{\alpha\beta}(r)$ between species $\alpha$ and $\beta$ (O, Si, or Na) which are given by~\cite{binder_kob_2011}: 

\begin{equation}
g_{\alpha\beta}(r)=\frac{V}{N_{\alpha}(N_{\beta}-\delta_{\alpha\beta})} \frac{1}{4\pi r^2}  \sum_{i=1}^{N_\alpha}\sum_{j=1, j \neq i}^{N_\beta}\langle\delta(r-|\vec{r}_i-\vec{r}_j|) \rangle,
\label{eq_2}
\end{equation}

\noindent
where $\langle \cdot \rangle$ denotes the thermal average, $V$ is the volume of the simulation box, $N_\alpha$ is the number of atoms of specie $\alpha$, and $\delta_{\alpha\beta}$ is the Kronecker-$\delta$. The coordination number $Z_{\alpha\beta}$ can then be calculated by integrating $g_{\alpha\beta}(r)$ in spherical coordinates up to its first minimum $r'_{\alpha\beta}$, i.e., 

\begin{equation}
Z_{\alpha\beta} = 4\pi \rho_\beta\int_{0}^{r'_{\alpha\beta}}g_{\alpha\beta}(r)r^2dr \quad,
\end{equation}

\noindent
where $\rho_\beta$ is the bulk number density of species $\beta$. The study of $g_{\alpha\beta}(r)$ shows that the cutoff distance $r'_{\alpha\beta}$ for the SiO pair can be fixed at 2.0~\AA, whereas for the SiNa and NaNa pairs the first minimum in $g_{\alpha\beta}(r)$ depends on temperature because of the thermal expansion of the system, but only weakly on composition. From the partial radial distribution functions, one can obtain pair correlation functions that can be directly compared with data from diffraction experiments. The total neutron pair distribution function $g_{\rm N}(r)$ is given by the expression:

\begin{equation}
g_{\rm N}(r)=\biggl(\sum_{\alpha,\beta}c_\alpha b_\alpha c_\beta b_\beta \biggr)^{-1}\sum_{\alpha,\beta}c_\alpha b_\alpha c_\beta b_\beta g_{\alpha\beta}(r),
\end{equation}

\noindent
where $c_\alpha$ is the fraction of $\alpha$ atoms and $b_\alpha$ is the neutron scattering length of the species, given by 5.803 fm, 4.1491 fm, and 3.63 fm for O, Si, and Na, respectively~\cite{sears1992neutron}. The total neutron structure factor $S_{\rm N}(q)$ can then be calculated via the space Fourier transform of $g_{\rm N}(r)$, i.e.,

\begin{equation}
S_{\rm N}(q)=1+4\pi\rho_0 \int_{0}^{\infty} \frac{{\rm sin}(qr)}{qr} [g_{\rm N}(r)-1] r^2 dr. 
\label{eq5}
\end{equation}

\noindent
To further understand the structural features, it is also useful to look at the partial structure factors 

\begin{equation}
S_{\alpha\beta}(q)=\delta_{\alpha\beta}+4\pi\rho_0 \bigl(c_\alpha c_\beta\bigr)^{1/2} \int_{0}^{\infty} \frac{{\rm sin}(qr)}{qr} [g_{\alpha\beta}(r)-1] r^2 dr.
\label{eq:s_alphabeta}
\end{equation}
 
Note that we use here the Ashcroft-Langreth expression instead of the conventional Faber-Ziman formalism~\cite{faber1965theory} because it better reflects the compositional effects on the pair correlation functions.

In addition to the standard two-point correlation functions, we also characterize the structure in 3D by using a set of novel four-point correlation functions~\cite{zhang2020pnas}. Specifically, we first construct a local coordinate system using a Si atom (particle 1) and any two of its nearest neighbor O atoms (particles 2 and 3): We define the position of particle 1 as the origin, the direction from particles 1 to 2 as the $z-$axis, and the plane containing the three particles as the $z-x$ plane. This local reference frame allows to introduce a spherical coordinate system $(\theta,\phi,r)$, and to measure the probability of finding any other particle at a given point in space, i.e., to measure a four-point correlation function (see Ref.~\cite{zhang2020pnas} for details). Note that this coordinate system can be defined for all triplets of neighboring particles, and these spatial density distributions can be averaged to improve the statistics.  Since this coordinate system is adapted to the local arrangement of the three particles, it allows to detect angular correlations in space that are not visible in $g(r)$ or other standard structural observables that depend only on two points. 

The so-obtained spatial distribution of the particles, $\rho(\theta,\phi,r)$, can be analyzed in a quantitative manner by decomposing it into spherical harmonics $Y_l^m$,

\begin{equation}
\rho(\theta,\phi,r) = \sum_{l=0}^\infty \sum_{m=-l}^{l}\rho_l^m(r) Y_l^m(\theta,\phi),
\label{eq7}
\end{equation}

\noindent
where the expansion coefficients $\rho_l^m$ are given by

\begin{equation}
\rho_l^m (r)=\int_0^{2\pi} d\phi \int_0^\pi d\theta \sin \theta \rho(\theta,\phi,r) Y_l^{m*}(\theta,\phi) \quad.
\label{eq8}
\end{equation}

Here $Y_l^{m*}$ is the complex conjugate of the spherical harmonic function of degree $l$ and order $m$~\cite{press_numerical_2007}.  In practice, the integral was carried out by sampling the integrand at a given radius $r$ over up to $10^8$ points using a shell of width 1.0~\AA. The strength of the 3D order can then be characterized by the square root of the angular power spectrum,  

\begin{equation} 
S_\rho(l,r)=\Bigl[(2l+1)^{-1}\sum_{m=-l}^{l}|\rho_l^m(r)|^2\Bigr]^{1/2}.
\label{eq9}
\end{equation}

For the NS$x$ system, the component with $l = 3$ is the most prominent one (see below) since the structural order is dominated by a tetrahedral symmetry. In the following we will see that in general
$S_\rho(l,r)$ exhibit an exponential decay at intermediate and large distances. This decay is related to the fact that $S_\rho(l,r)$ is sensitive not only to the angular dependence of the density distribution, i.e., the symmetry of the field $\rho(\theta,\phi,r)$, but also to the amplitude of the signal. In order to probe the symmetry aspect of the structural order at large distances, it is therefore useful to consider a \textit{normalized} density distribution $\eta(\theta,\phi,r)$, which is defined as

\begin{equation}
\eta(\theta,\phi,r)=\frac{\rho(\theta,\phi,r)-\rho_{\rm min}(r)}{\rho_{\rm
max}(r)-\rho_{\rm min}(r)} ,
\label{eq10}
\end{equation}

\noindent
where $\rho_{\rm max}(r)$ and $\rho_{\rm min}(r)$ are the maximum and minimum of $\rho(\theta,\phi,r)$, respectively. To get robust results for $\eta$, we used for $\rho_{\rm max}(r)$ and $\rho_{\rm min}(r)$ the mean of, respectively, the highest and lowest densities in a pixel covering 1\% of the sphere surface. Below we will see that the square root of the angular power spectrum of $\eta(\theta,\phi,r)$, $S_{\eta}(l,r)$, allows to detect feature of the orientational order of the structure that are invisible in the standard two point correlation functions.

\section{Results and discussion}
\subsection{Validation of the liquid and glass structures}

\begin{figure*}[t]
\center
\includegraphics[width=0.67\textwidth]{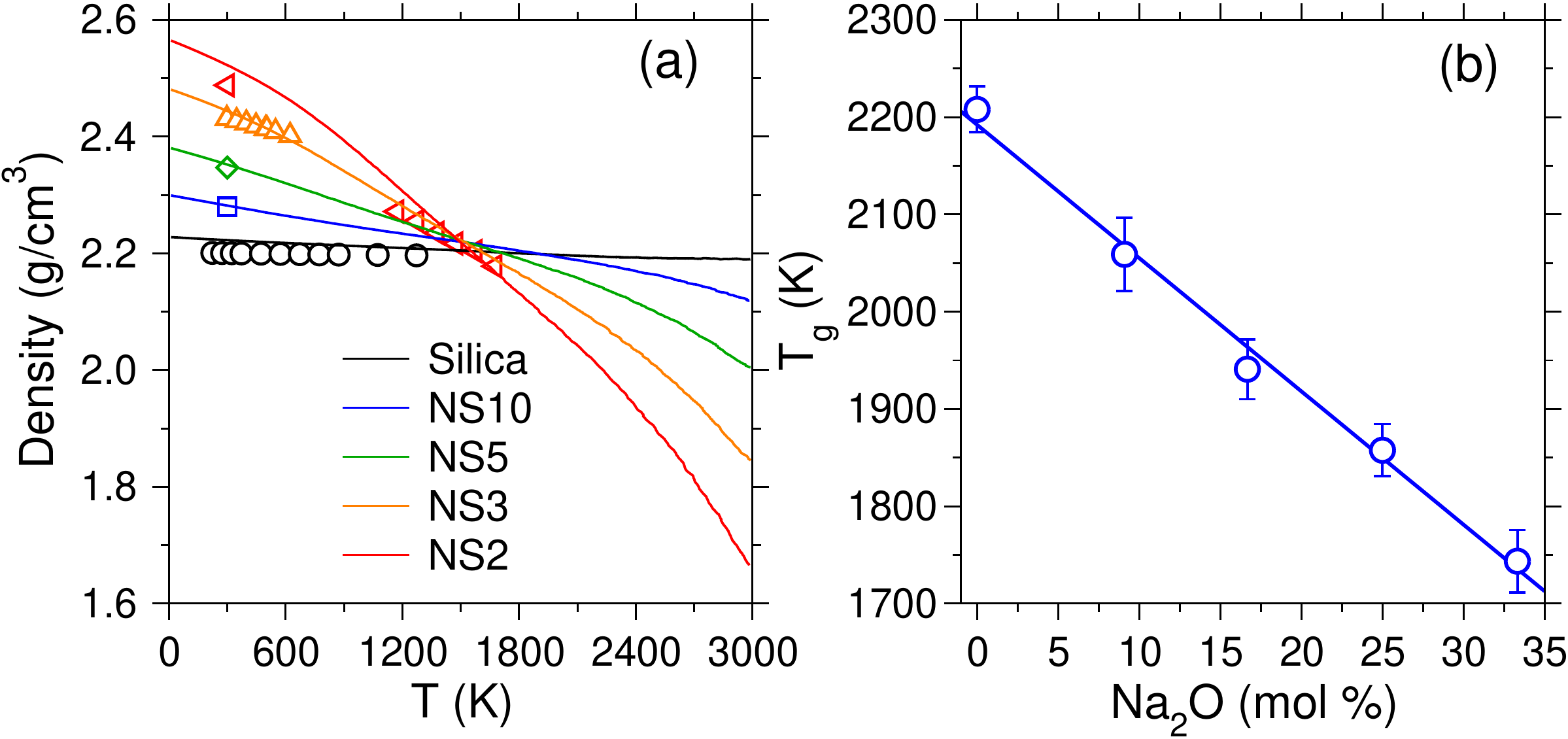}
\includegraphics[width=0.32\textwidth]{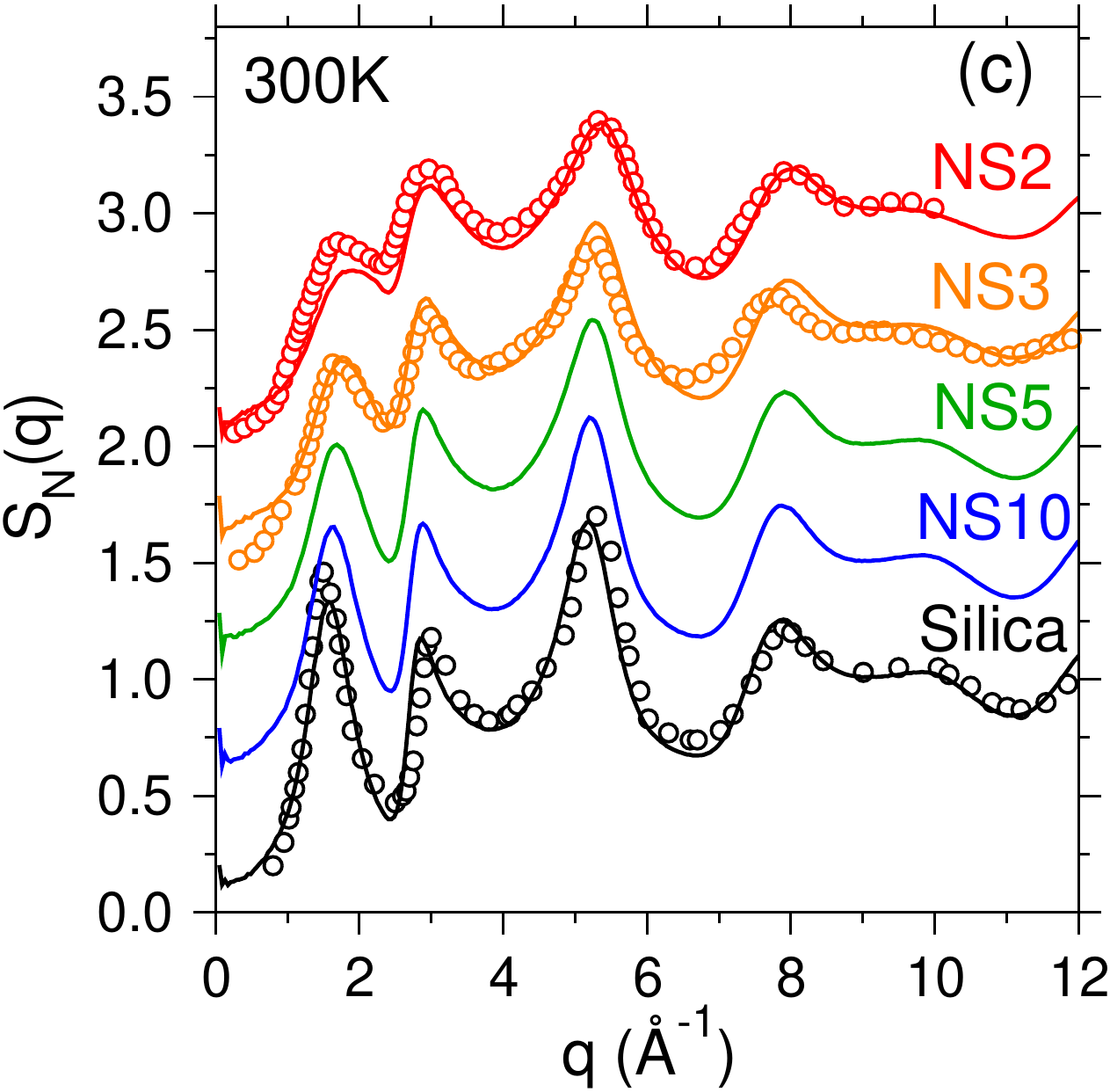} 
\caption{(a) Temperature dependence of the mass density of the silica and NS$x$ systems. Also included is experimental data~\cite{shartsis1952density,bansal_handbook_1986} at various temperatures (symbols with the same color as the simulation data). (b) The glass transition temperature as estimated from the intersection point of two linear fits to the low- and high-temperature data of the enthalpy. The solid line is a liner fit to the data points. (c) Total neutron structure factor at 300~K. Symbols are experimental data for silica, NS3, and NS2 which are taken from Refs.~\cite{susman_temperature_1991,pohlmann_strcuture_2005,misawa1980short}. The curves are shifted vertically by multiples of 0.5 to improve readability.
}
    \label{fig_rho_Tg}
\end{figure*}

We first present in Fig~\ref{fig_rho_Tg}(a) the temperature-dependence of the mass density of the NS$x$ systems having various Na$_2$O concentrations (solid curves). One recognizes that the simulations reproduce well the compositional and temperature dependence of the experimental values of the densities (symbols)~\cite{shartsis1952density,bansal_handbook_1986}. 
This indicates that the atomic structure of the simulated samples reflects reasonably well the one of real sodo-silicate glass-formers. We observe that the density of the liquid at high temperatures decreases considerably with increasing Na$_2$O concentration, whereas at low $T$'s the glass density shows the opposite dependence on composition. 
That an increasing Na concentration gives rise to a stronger $T-$dependence of density (and thus of the thermal expansion coefficient) can be rationalized by the change of structure when Na$_2$O is added to the SiO$_2$ system: Na is a typical network modifying species which depolymerizes the Si-O network by breaking the connection between [SiO$_4$] tetrahedra and hence the structure becomes more flexible, i.e., the expansion coefficient increases. Surprisingly we observe that the curves for the different systems cross at around 1500~K, i.e., all the systems have basically the same density. Although we are not aware of any experimental data that allows to check whether real systems show the same feature, this result should be taken as motivation to carry out experiments that investigate this point in more detail. We also mention that the presented $T-$ and $f-$dependence of the mass density is very similar to the one of the particle-(atom) density, a result that is reasonable since the mass of a Na$_2$O unit happens to be very close to the one of SiO$_2$.

Approximating at high and low temperatures the $T-$dependence of the enthalpy by two straight lines (not shown), one can estimate from their intersection point the glass transition temperature $T_g$~\cite{vollmayr_cooling-rate_1996}. Panel (b) shows that $T_{\rm g}$ exhibits basically a linear dependence on the concentration of Na. Note that these values are considerably higher than the experimental values (which are around 1500~K for silica and 750~K for NS2) as obtained, e.g., from calorimetric measurements~\cite{knoche_non-linear_1994}, i.e., a technique in which the sample is cooled significantly slower than the quench rates used here. 

\begin{figure*}[ht]
\center
    \includegraphics[width=0.9\textwidth]{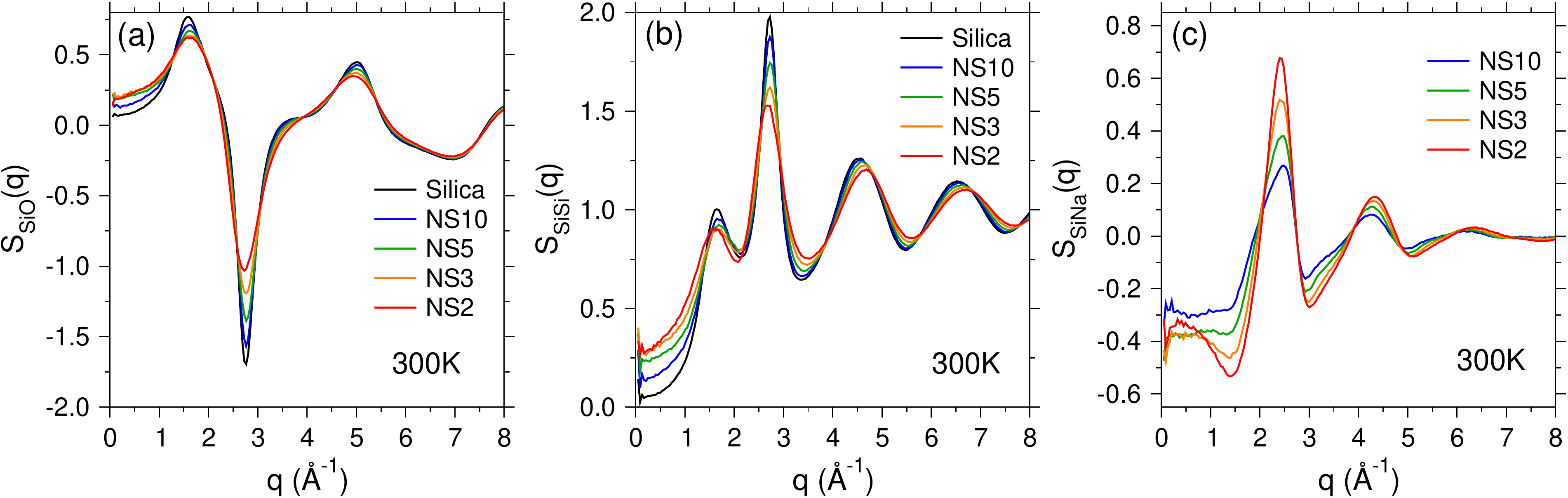}
    \includegraphics[width=0.9\textwidth]{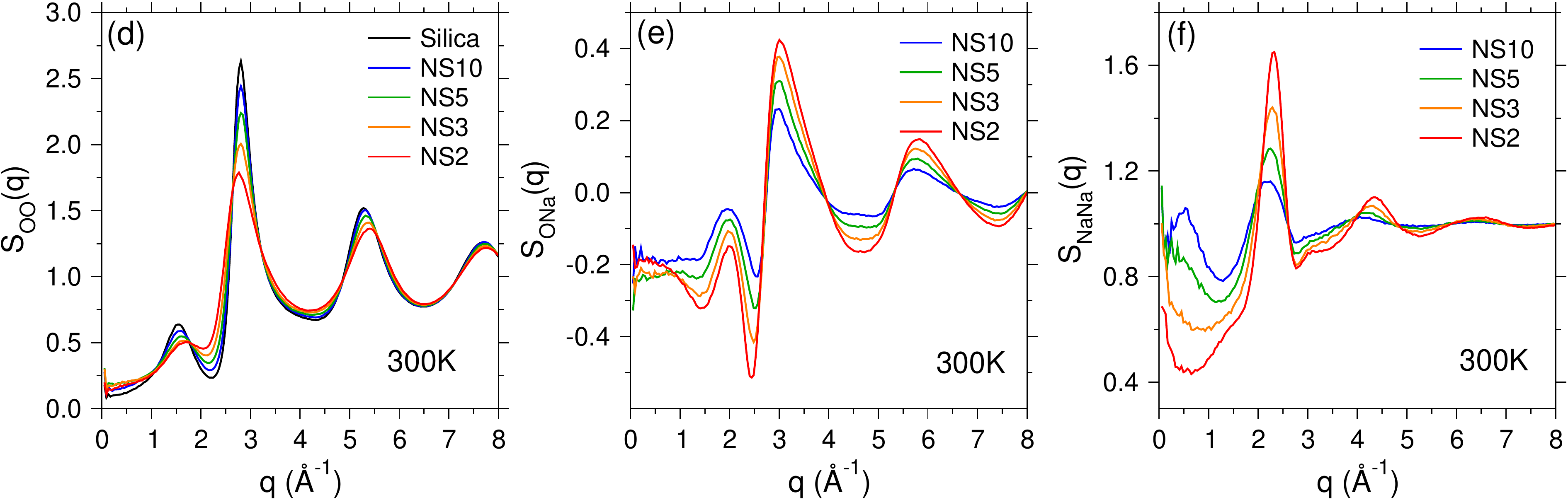}
    \caption{The six partial structure factors at 300~K, obtained using the Ashcroft-Langreth formalism, see Eq.~(\ref{eq:s_alphabeta}). 
    }
    \label{fig_sq-partial-nsx-300K}
\end{figure*}
Finally, we compare in Fig.~\ref{fig_rho_Tg}(c) the total neutron structure factor, $S_{\rm N}(q)$from Eq.~(\ref{eq5}), calculated for the simulated NS$x$ glasses at 300~K with the ones measured in experiments. 
Overall, one sees that our simulation results are in excellent agreement with the experimental data, for the whole composition range investigated. It is worthwhile to note that the interaction potential used in this study was parameterized using the reference data from \textit{ab initio} calculations (for liquids) and experiments (for glasses) only for silica~\cite{sundararaman_new_2018} and NS4~\cite{sundararaman_new_2019}. The good agreement between our simulations and experiments across a wide composition range indicates thus the good transferability of this effective potential. It can thus be expected that the present simulations reproduce at least semi-quantitatively the structural features of this type of glass-former. In the following, we will hence present a comprehensive account of the structure of the liquids and glasses, notably their medium range order.

\begin{figure*}[ht]
\center
\includegraphics[width=0.90\textwidth]{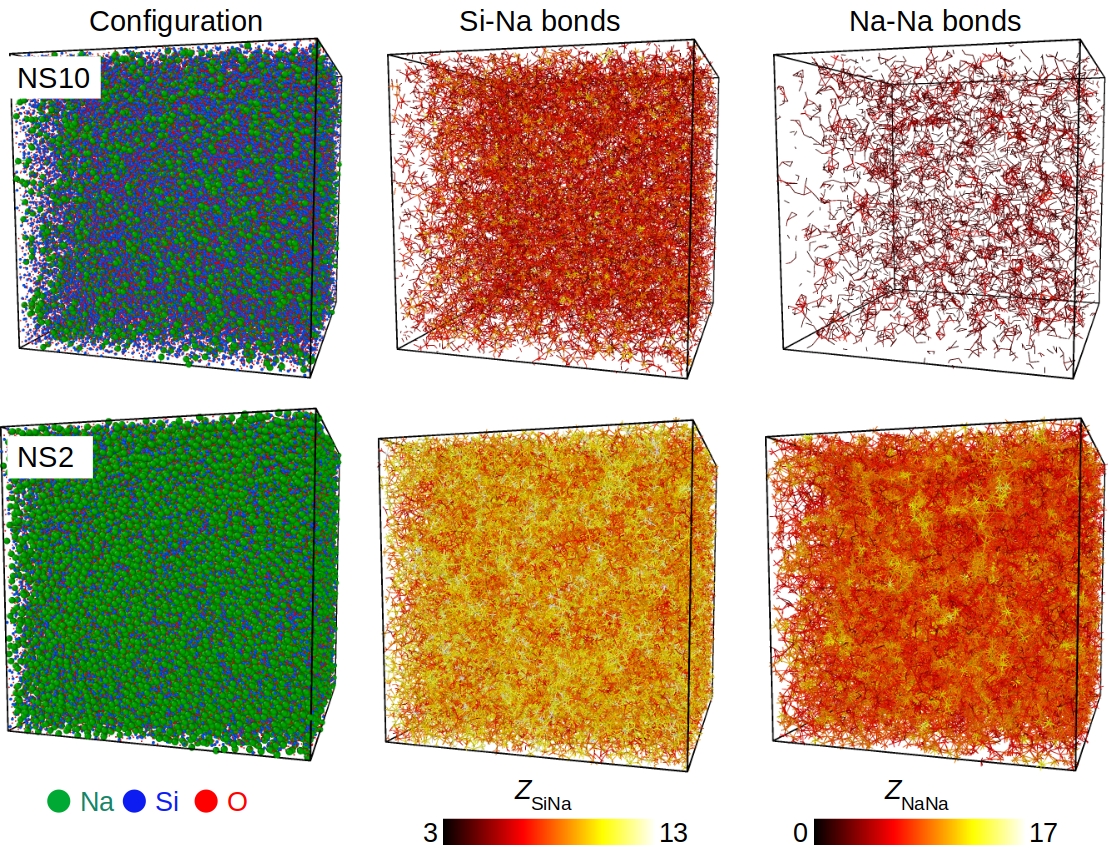}
\caption{From left to right: Snapshots showing the atomic configuration, Si-Na bonds, and Na-Na bonds, respectively. Upper and lower panels are for the NS10 and NS2 glasses, respectively. Bonds are displayed if the distance between two atoms is smaller than the first minimum of their corresponding $g_{\alpha\beta}(r)$. The color of the bonds are given by the coordination number of the corresponding atom. The color scheme is given at the bottom of the middle and right column.  
The coordination number is obtained by counting the number of bonds. Oxygen atoms are shown small to make the spatial distribution of Na and Si clearer.
}
    \label{fig_snapshots-config-bonds-nsx-300K}
\end{figure*}

\subsection{Spatial distribution of sodium}
We first probe the spatial distribution of the network-modifying Na atoms using standard two-point structural observables. Figure~\ref{fig_sq-partial-nsx-300K} shows the partial structure factors $S_{\alpha\beta}(q)$, see Eq.~(\ref{eq:s_alphabeta}), for the glasses at 300~K. Overall, one observes that all partials show a smooth dependence on Na concentration and that some interesting features emerge as the composition is varied. Notably $S_{\rm SiNa}(q)$ and $S_{\rm ONa}(q)$ show a  broad peak at $q<1$~\AA$^{-1}$ at the highest Na concentrations [panels (c) and (e)], while $S_{\rm NaNa}(q)$, panel (f), shows a peak at small $q$ which gradually moves to smaller wave-vectors with increasing Na concentration. 
These results indicate that the spatial distribution of the Na atoms on intermediate and large length scales depends significantly on the composition, although we have not seen any sign of a mesoscopic phase separation which would be indicated by a strong increase of the signal at small $q$~\cite{zhang2020potential} (see also the snapshots in Fig.~\ref{fig_snapshots-config-bonds-nsx-300K}).

To further understand the spatial distribution of the atoms (in particular Na), we show in Fig.~\ref{fig_snapshots-config-bonds-nsx-300K} for NS10 and NS2 the atomic configurations at 300~K. One notices that for both glasses the atoms of different types are seemingly mixed quite homogeneously, left panels. However, a closer inspection of the configurations in terms of the SiNa and NaNa bonds (middle and right panels, respectively) shows that there is notably inhomogeneity in the spatial distribution of Na. Specifically, the Na atoms are forming an irregular structure of blobs (or ``pockets'' as referred in previous studies~\cite{jund2001channel}) that are connected by filaments. A closer inspection reveals that these Na pockets percolate with increasing Na concentration, eventually leading to the formation of a network of Na channels. We thus conject that the rise of the signal at small $q$ is the signature of the Na diffusion channels as reported in previous studies~\cite{ingram1989ionic,horbach2002channels,meyer_channel_2004}. More quantitative results regarding the local environment of Na will be presented below.

\begin{figure}[th]
\center
\includegraphics[width=0.68\textwidth]{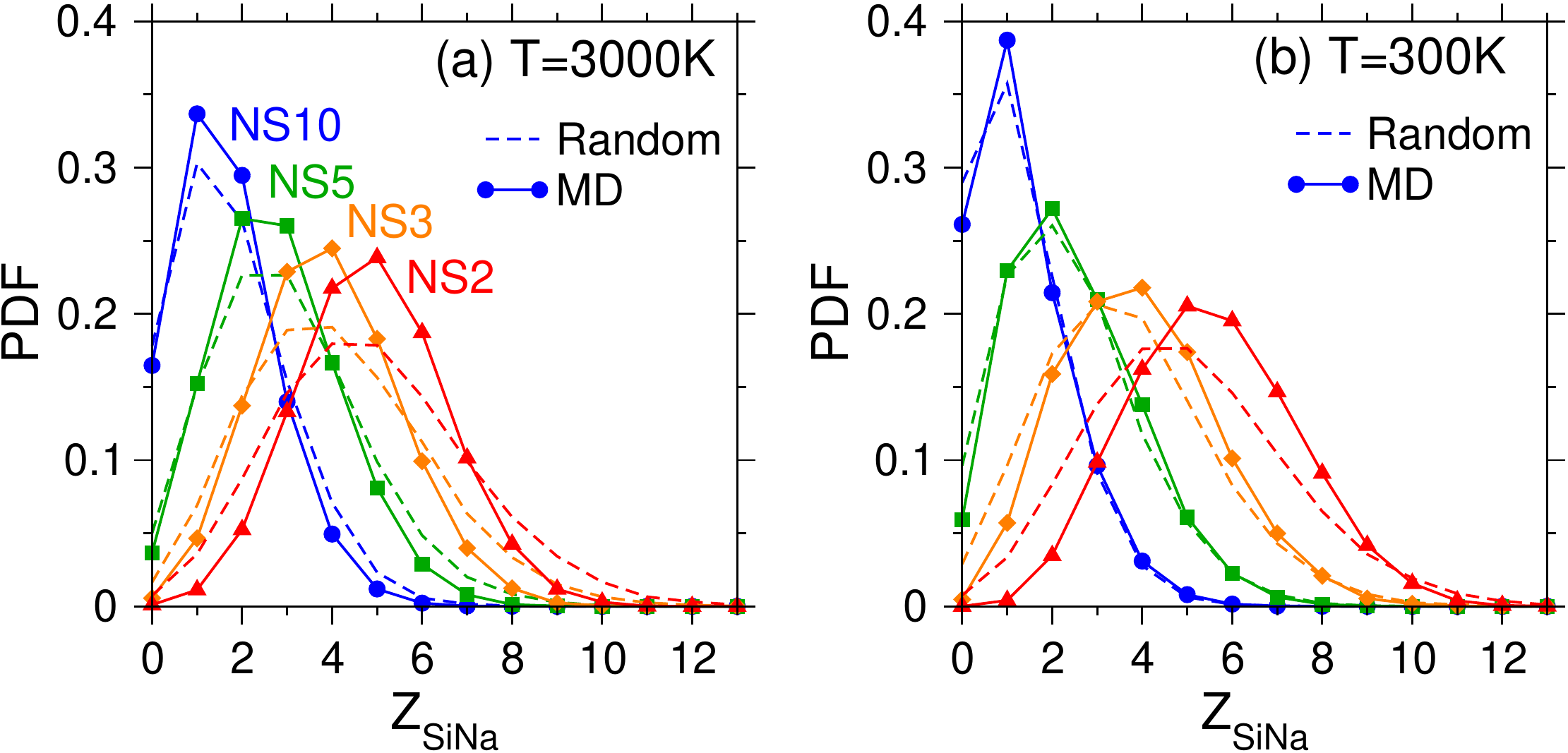}
\includegraphics[width=0.68\textwidth]{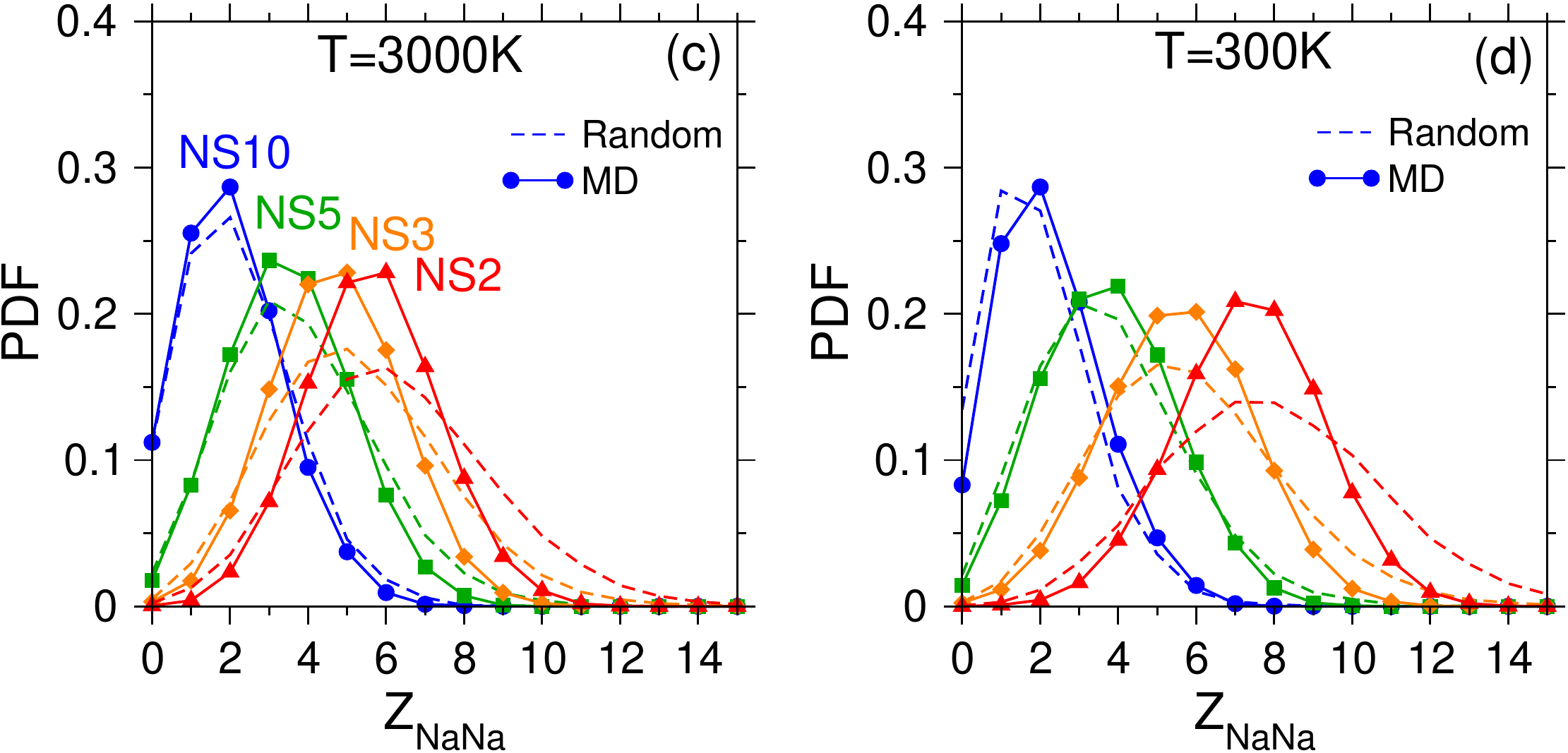}
\caption{Coordination number, $Z_{\rm XNa}$, i.e., the number of Na in the nearest neighborhood of X.	(a)-(d): Distribution of $Z$ for the liquids at 3000~K and the glasses at 300~K. 
For SiNa, the cutoff distance for counting the neighbors corresponds to the first minimum of $g_{\rm{SiNa}}(r)$, which for the sample at 3000~K and 300~K is, respectively, at 4.75~\AA\ and 4.15~\AA. For Na-Na, the corresponding cutoffs are 5.15 and 4.85~\AA\ for 3000~K and 300~K, respectively.
}
\label{fig_cn-nsx-sina-nana}
\end{figure}

Figure~\ref{fig_cn-nsx-sina-nana} presents the distribution of the coordination number of Na around a Si (${Z}_{\rm SiNa} $) and a Na ($ {Z}_{\rm NaNa} $) at high and low temperaures. 
For the sake of comparison we include also the data (using the same $T$-dependent cut-off distance to define a nearest neighbor as in the simulated structure) for configurations with the same composition and particle density, but with a completely random distribution of atoms, i.e., in this case the radial distribution function is featureless. 
For low Na concentration, NS10, one recognizes from these distributions that the local environment of Na is close to random, panels (a-d). With increasing Na concentration the difference between the simulation result and the random curve becomes larger, indicating that the local environment of Na becomes more ordered. It is, however, remarkable that the location of the peak for the real distribution coincides very well with the random one, and this holds even at high Na concentration. Hence, one concludes that the position of the peak is determined solely by the particle density and it is mainly the width of the distribution that reflects the ordering of the structure in that the real distribution is narrower. Comparing the distributions for high $T$'s with the ones at low $T$'s, one notices that they are surprisingly similar. In the following we study this point in more detail.

\begin{figure*}[th]
\center
\includegraphics[width=0.85\textwidth]{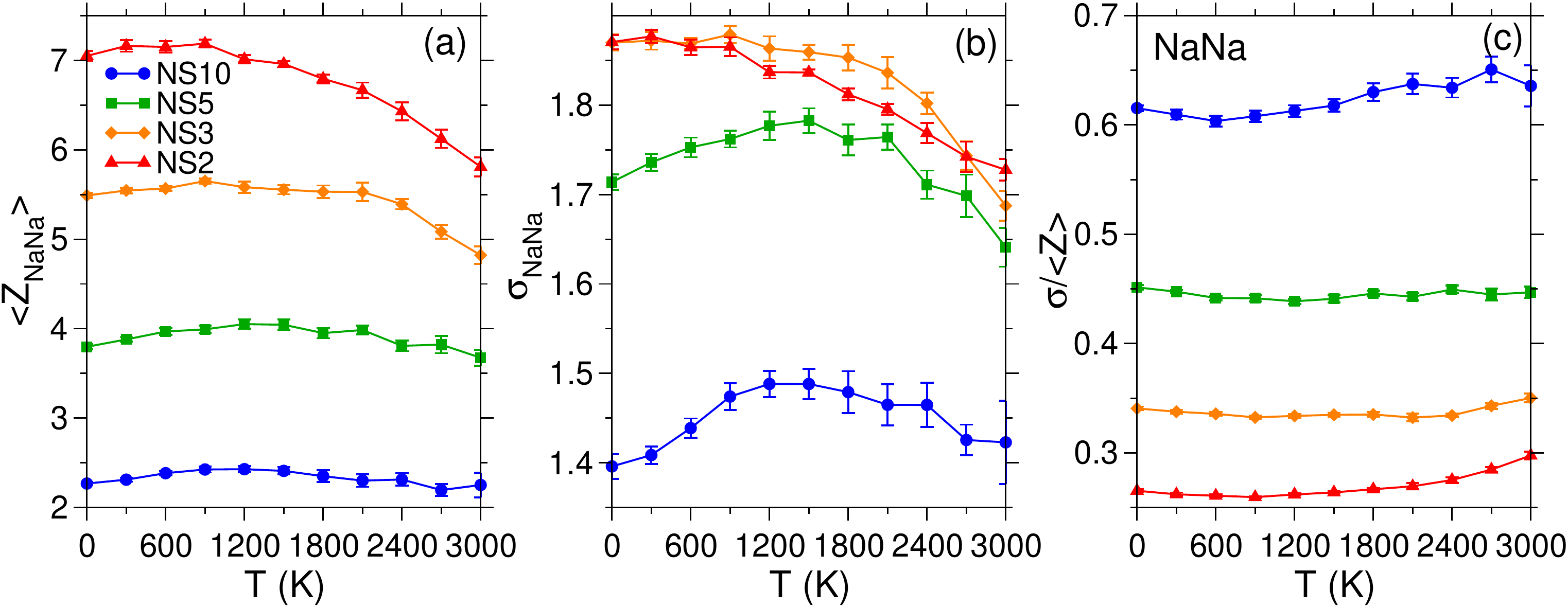}
\includegraphics[width=0.85\textwidth]{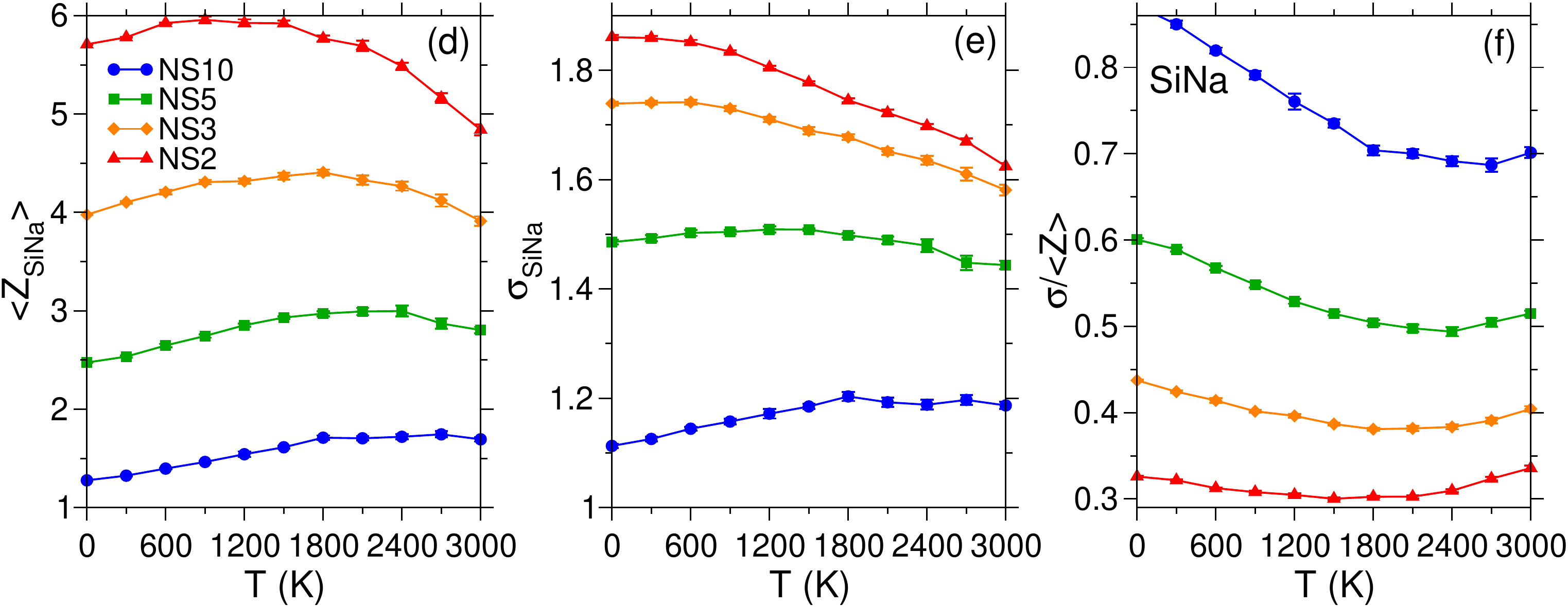}
\caption{From left to right: Temperature-dependence of the mean coordination number $\langle Z \rangle$, the standard deviation $\sigma$ of the coordination number distribution, and the relative width $\sigma/ \langle Z \rangle$ of the coordination number distribution. The upper (a-c) and lower (d-f) panels are for the NaNa and SiNa coordinations, respectively.
}
\label{fig_cn-nsx-sina-nana-Tdepend}
\end{figure*}

In order to elucidate the $T-$dependence of the local environment of Na, we have calculated the mean coordination numbers (CN) $\langle {Z}_{\rm NaNa} \rangle$ and $\langle {Z}_{\rm SiNa} \rangle$ and the standard deviation of their distributions, see Fig.~\ref{fig_cn-nsx-sina-nana-Tdepend}. From panels (a) and (d), one  recognizes that the $T-$dependence of $\langle {Z}_{\rm NaNa} \rangle$ and $\langle {Z}_{\rm SiNa} \rangle$ becomes more pronounced with increasing Na concentration. The Na-rich systems (i.e., NS5-NS2) display a maximum at intermediate $T$'s, the location of which depends on the Na concentration and the correlation (SiNa or NaNa). The observed non-monotonic $T-$dependence of the CNs hints that the local structure of the Na atoms is influenced by competing mechanisms, the strength of which is $T-$dependent, and below we will find further evidence for this.  Panels (b) and (e) present the standard deviation $\sigma$ of the two distributions and one recognizes that this quantity shows a $T-$dependence that is qualitatively similar to the one of the CNs. However, the $f-$dependence of the two quantities is quite different, which indicates that the relative width of the distribution has a non-trivial $f$-dependence. We therefore present in panels (c) and (f) the ratio $\sigma/\langle Z\rangle$ which serves as an indicator for the heterogeneity of the Na neighborhoods. 
A decrease of $\sigma/\langle Z \rangle$ implies that the environment of the atom becomes more uniform. For SiNa one sees that this is indeed the case if $T$ is lowered from high $T$ to intermediate ones, a behavior that is expected since decreasing temperature should make the sample more homogeneous. However, if temperature is lowered even further, one finds that the ratio starts to increase significantly, indicating that the Na environment around a Si atom becomes again more heterogeneous. This rise is modest for systems with high Na concentration, but quite pronounced for systems with low $f$. 
Note that the temperature at which this crossover is observed is close to $T_g$ and hence the change of $T-$dependence might be related to out-of-equilibrium effects. However, below we will see that this is not the case.
In comparison, the $T-$dependence of $\sigma/\langle {Z} \rangle$ for NaNa is less pronounced than that for SiNa, but also here one finds a change of the $T-$dependence of $\sigma/\langle {Z} \rangle$ at intermediate $T$'s. Below, we will see that these observations on the Na-related SRO can be linked to the non-trivial $T-$dependence of the MRO.

\subsection{Ring structures}

Although the presented $T$- and $f$-dependence of the coordination numbers are instructive to characterize the short range order, they do not allow to gain insight into the structure of the system at larger length scales (say $r>5$~\AA). For this it is thus useful to consider quantities that are sensitive on these scales, such as the ring statistics discussed now. 

\begin{figure}[ht]
\center
    \includegraphics[width=0.68\textwidth]{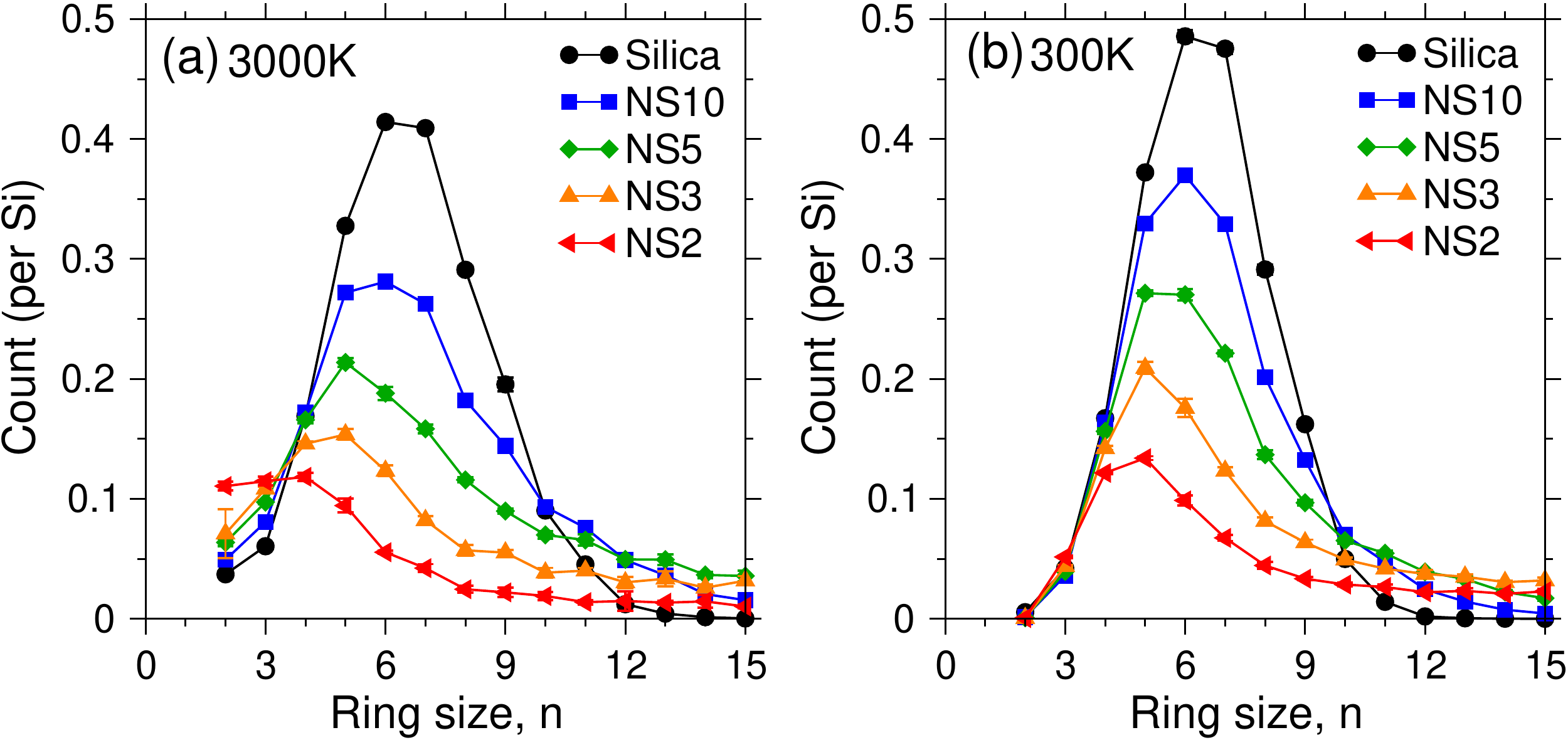}
    \includegraphics[width=0.68\textwidth]{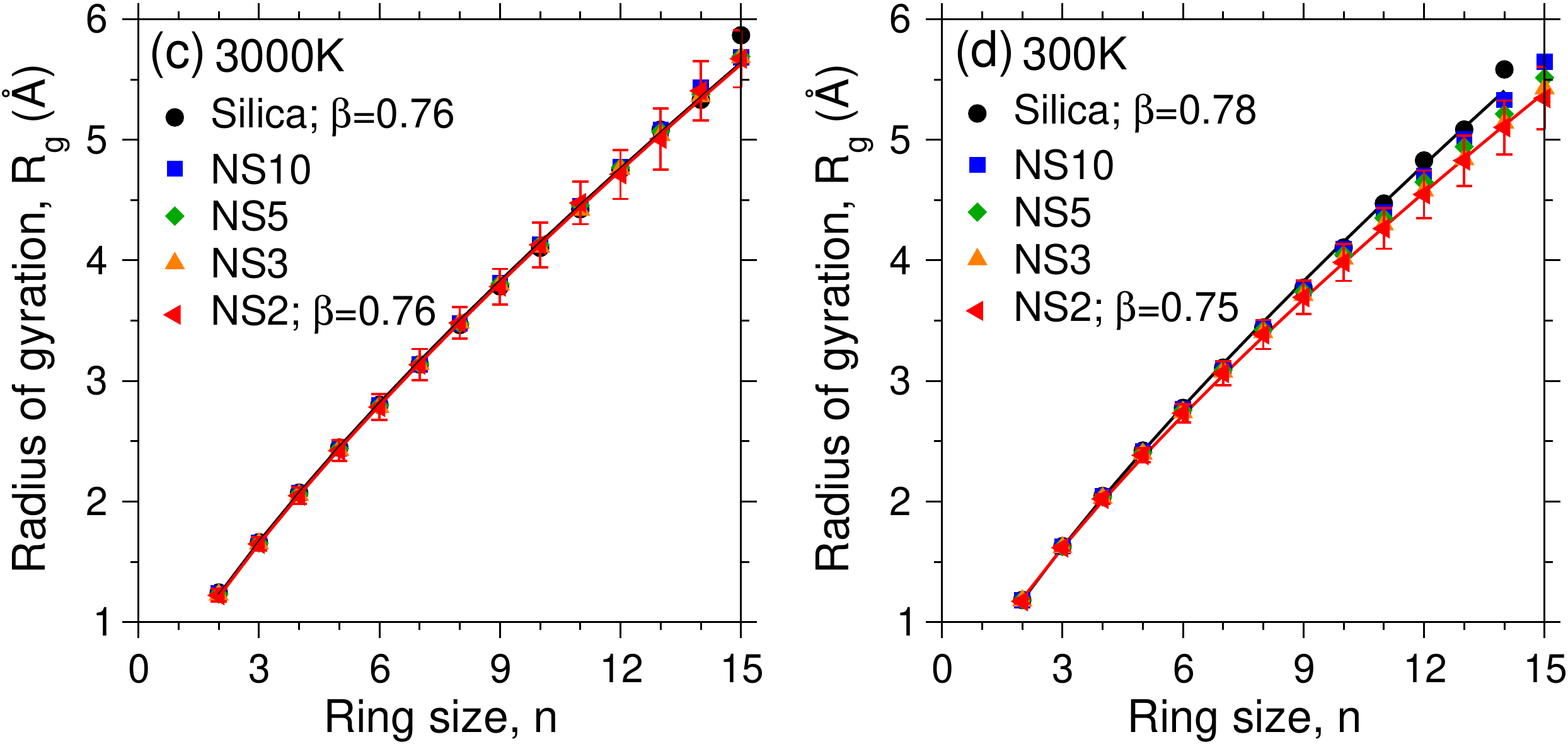}
    \caption{Ring statistics for the liquids at 3000~K and the glasses at 300~K. (a) and (b): The number of rings normalized with respect to the number of nodes (i.e. Si atoms). 
    (c) and (d): Radius of gyration of the rings of various sizes. Vertical bars represent the standard deviation of the $R_{\rm g}$ distribution for the NS2 glass. (Note that this width depends only weakly on composition).
    Solid lines are power-law fits to the data of silica and NS2, and $\beta$ is the exponent. 
    }
        \label{fig_rings-nsx} 
\end{figure}

We define the ring size $n$ as the number of Si atoms in the smallest closed loop of Si and O atoms, the so-called primitive rings~\cite{yuan2002ring,leRoux2010ring}, i.e., none of the identified rings can be decomposed into two smaller rings. Figure~\ref{fig_rings-nsx} displays the PDF of the ring size for the different systems at high and low temperatures, panels (a) and (b), respectively. 
One observes that for the liquids at 3000~K, an increasing Na concentration favors the formation of small rings, notably two-membered rings formed by two [SiO]$_4$ tetrahedra sharing an edge, panel (a), while at low $T$'s the concentration of small rings is low and for these $n$'s the distribution is independent of Na concentration, panel (b). This result is reasonable since small rings are energetically unfavourable and hence are avoided at low $T$. For high Na concentrations, one finds at large $n$ a pronounced tail in the ring size distribution, which reflects the depolymerization of the Si-O network, making that some of the rings become very large since cutting a ring will inevitably result in the creation of a ring with larger $n$. Qualitatively similar results for the ring statistics were reported in previous MD studies of sodium silicate glasses~\cite{du_medium_2004,sundararaman_new_2019}.

The spatial extent of these rings can be characterized by their radius of gyration, $R_{\rm g}$, given by the expression
\begin{equation}
R_{\rm g}^2 = \frac{1}{2n}\sum_{i=1}^{2n} ({\vec{r}_i} - \vec{R}_{\rm C})^2 \quad,
\end{equation}
where $\vec{R}_{\rm C}$ is the location of the geometrical center of the ring. 
(Note that since a ring contains equal number of Si and O atoms, the total number of atoms constituting a $n$-membered ring is 2$n$.)  
Panels (c) and (d) show that $R_{\rm g}(n)$ is described very well by a
power-law $n^\beta$ with an exponent $\beta$ that is about 25\% below the trivial value of 1.0, which indicates that with increasing size rings become more and more crumbled, although to a first approximation they can be considered as planar (which corresponds to an exponent of 1.0).
It is interesting to note that at high $T$ the exponent is independent of the Na concentration while at low $T$ there is a weak but noticeable dependence in that an increasing Na concentration gives rise to rings that are slightly more crumbled (see values of $\beta$ in the legend). Note that not only the exponent of the power-law is independent of the Na concentration, but also the pre-factor. 
This means that the geometry of the rings is invariant with a change of Na concentration and a comparison of the two panels shows that it is also only weakly dependent on $T$. Although this finding is not surprising for small rings because of geometric constraints, this property is remarkable for large $n$. Note that this result implies that the main effect of temperature is not that the geometry of the rings are changing, but only the probability to find a ring of a given length.
From the analysis of the rings, one recognizes that the most abundant rings, $n=5-7$, typically have a radius of 2.5-3.5~\AA, i.e. the atoms in such a ring are typically separated by a distance around 6~\AA. As we will see later, this intermediate length scale is related to the $r-$dependence of various correlation functions.

\subsection{Radial distribution functions}

\begin{figure*}[ht]
\center
     \includegraphics[width=16cm]{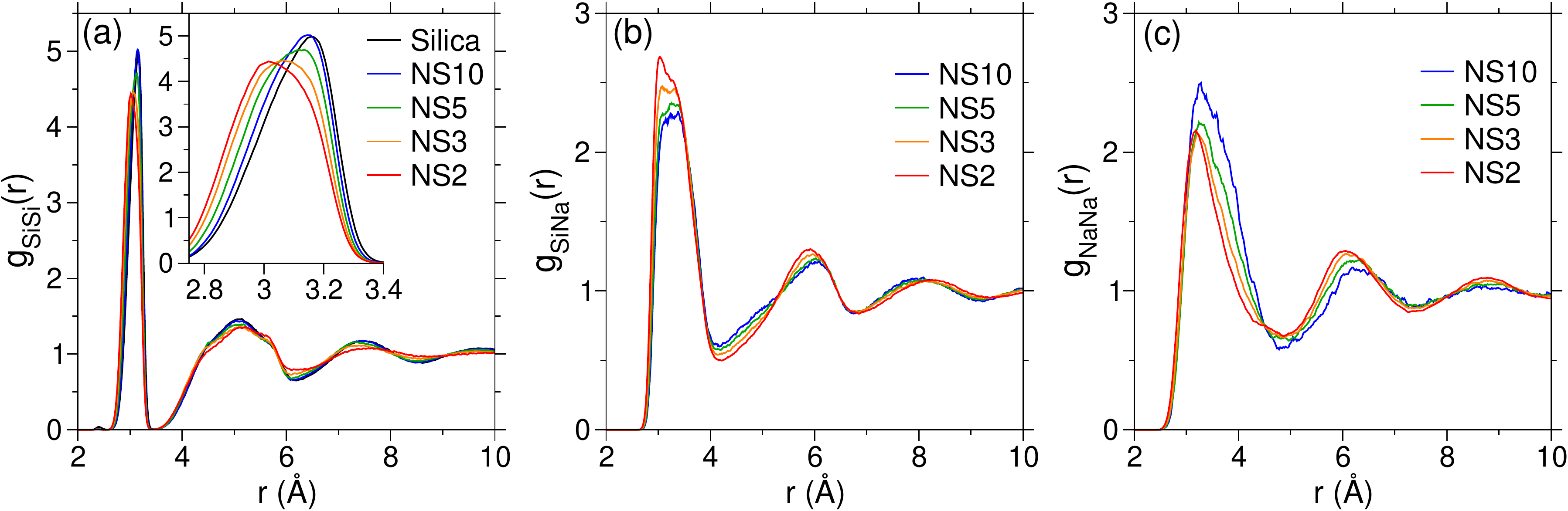}
    \caption{Partial radial distribution function at 300~K. (a)-(c) are for the SiSi, SiNa, and NaNa pairs, respectively. The inset in (a) is a zoom on the first peak.\\  
}
\label{fig_gr-nsx-sisis-nana-300K}
\end{figure*}

Figure~\ref{fig_gr-nsx-sisis-nana-300K} shows the partial radial distribution functions $g_{\alpha\beta}(r)$, defined in Eq.~(\ref{eq_2}), for the glasses at 300~K. For SiSi, panel (a), one sees that the position of the main peak remains fixed for low Na concentration while for NS3 and NS2 it has moved to smaller distances and the peak becomes broader, in agreement with the observed increase in density documented in Fig.~\ref{fig_rho_Tg}(a). For the SiNa and NaNa correlations, panels (b) and (c), the position of this peak gradually shifts to smaller distances with increasing Na concentration. 
Interestingly, one observes that in $g_{\rm SiNa}(r)$ a double peak emerges gradually as the Na concentration increases. A more detailed analysis of the various structural environments of Na (not shown) reveals that this double peak is related to the number of neighboring oxygen atoms of the Na atom: Having more oxygen neighbors results in a smaller Si-Na distance since the presence of the oxygen atoms gives rise to an attractive component in the effective Na-Si interaction.
A further observation is that the width as well as the height of the first peak in $g_{\rm NaNa}(r)$ decrease with increasing Na concentration, and that this width shows a stronger  dependence on composition than the height. This indicates that the local Na arrangement becomes more structured with increasing Na concentration and below we will show that this ordering can be attributed to the increasing dominance of energetic terms. 
Also, as we will see later in the context of the static structure factors, this result reflects the fact that at low Na concentrations, these atoms have the trend to form small zones with high Na concentration.
Apart from these observations, the partial radial distribution functions do not seem to allow to identify a MRO, at least not in the type of representation used in Fig.~\ref{fig_gr-nsx-sisis-nana-300K}. 

\begin{figure*}[th]
\center
\includegraphics[width=16cm]{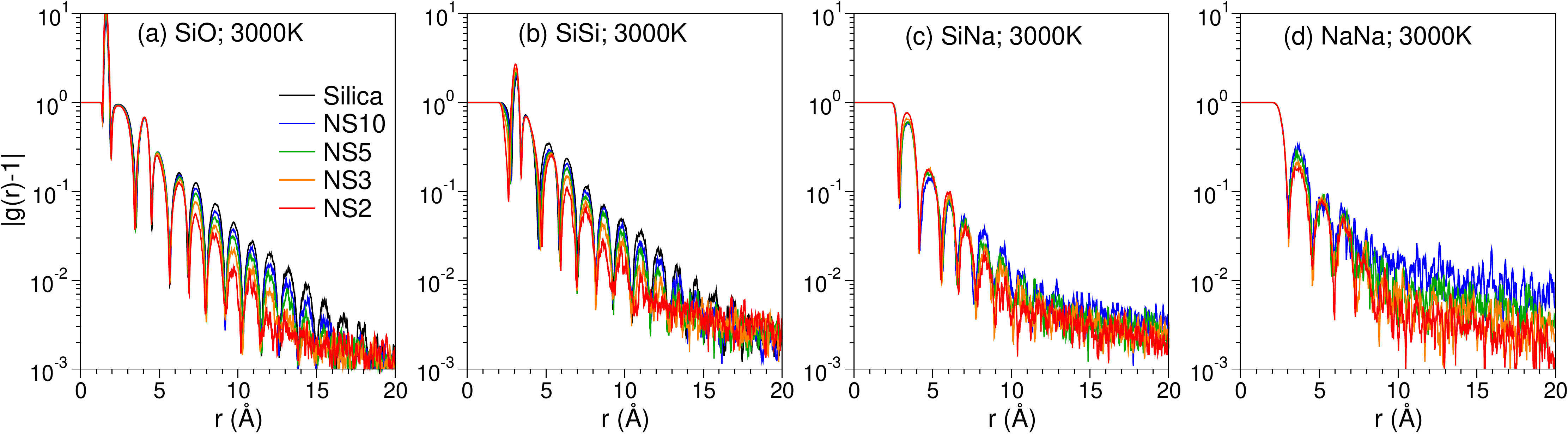}
\includegraphics[width=16cm]{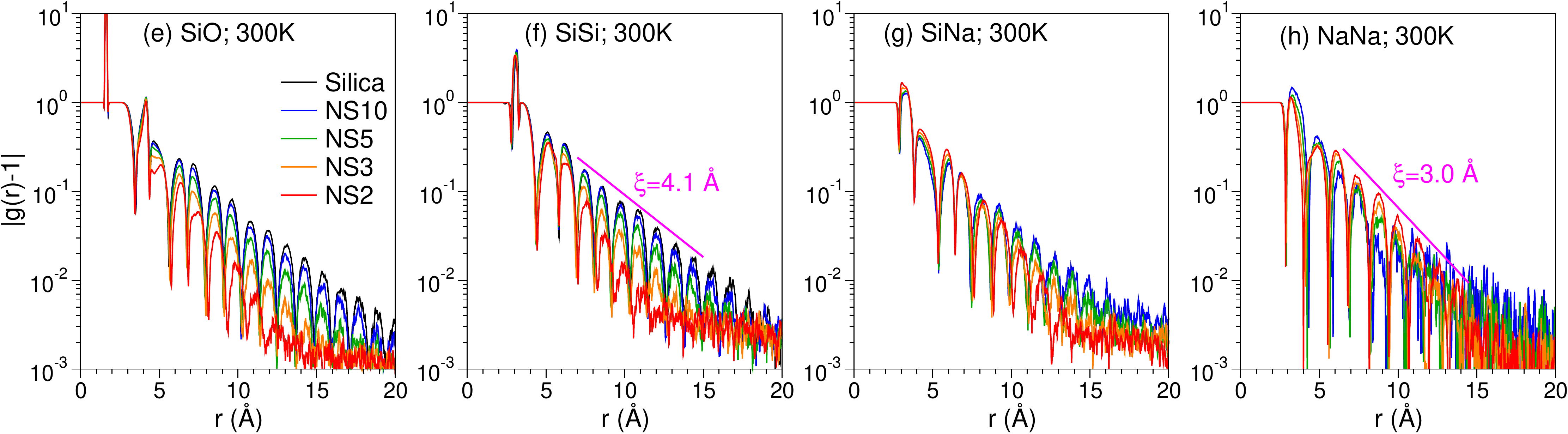}
\includegraphics[width=16cm]{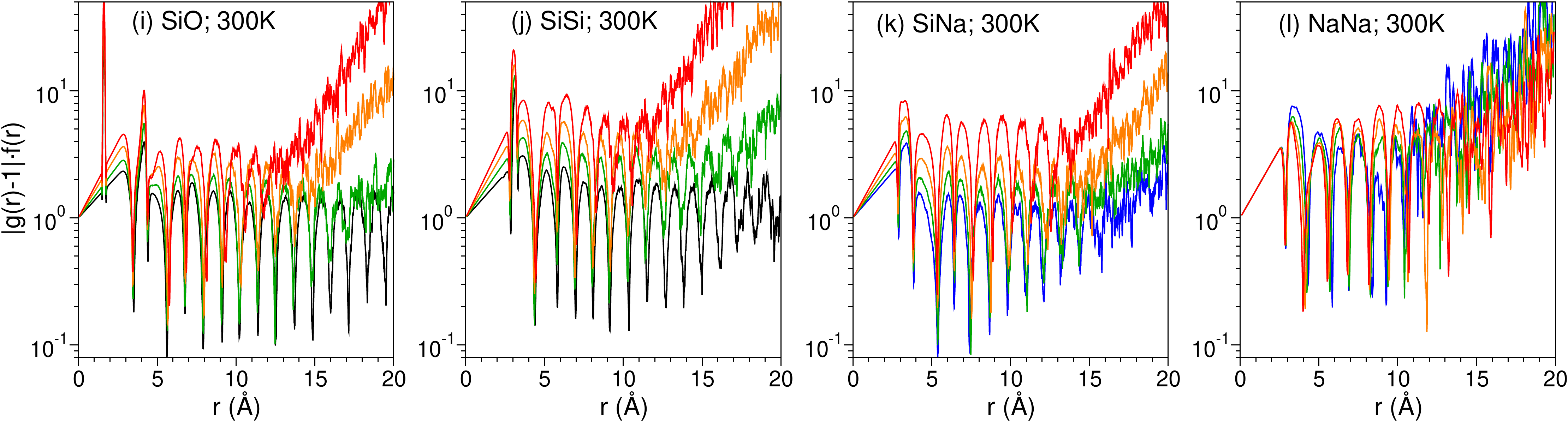}
\caption{$|g(r)-1|$. From left to right, the panels are for SiO, SiSi, SiNa, and NaNa correlations. (a)-(d): 3000~K. (e)-(h): $|g(r)-1|$ at 300~K. 
(i)-(l): $|g(r)-1| \cdot f(r)$, i.e., $|g(r)-1|$ at 300~K multiplied by an exponential $f(r)$ to make the curves more or less horizontal. Note that this exponential function depends on composition for the SiO, SiSi, and SiNa correlations, whereas it is independent of composition for the NaNa correlation. 
The pink solid lines in panels (f) and (h) are a guide to the eye to see the exponential decay of the function at large $r$.
}
\label{fig_gr-1_nsx-l03-SiX}
\end{figure*}

The MRO is expected to be noticeable at distances between 5 and 20~\AA, a range in which the $g_{\alpha \beta}(r)$'s shows only small deviations from unity. In order to better see the $r$-dependence of $g_{\alpha\beta}(r)$ at these distances, we therefore present in Fig.~\ref{fig_gr-1_nsx-l03-SiX} the function $|g_{\alpha\beta}(r)-1|$ for the various partials in the NS$x$ system. At high $T$, panels (a)-(d), the $|g_{\alpha\beta}(r)-1|$ decay at intermediate and large distances in an exponential manner and show little dependence on Na concentration, i.e., the medium range order that is present is not affected significantly by the Na concentration or is not detectable by these correlation functions. 
(Note that the flattening of the curves at large distances is due to a deteriorating statistics.) 
When the samples are in the glass state, panels (e)-(h), one recognizes that the complex dependence of the short range order on composition discussed in Fig.~\ref{fig_gr-nsx-sisis-nana-300K} 
extents up to distances around 7~\AA, i.e., to approximately the third nearest neighbor shell. 
For larger distances one finds again an exponential decay allowing to define a correlation length $\xi$ (see Eq.~\ref{eq_exp-decay-gr} below) which will be discussed below in more detail. 
This cross-over in the $r$-dependence indicates that the systems have at short distances a more complex order than at large distances. The rationale for this observation is that at short length scales there is a large variety of structural motifs and the concentration of each will depend strongly on temperature and composition thus affecting the local structure. For distances larger than around 7~\AA, which, as discussed above, corresponds to the size of the most frequent rings, the structural consequence of these diverse local environments is averaged out and gives rise to a new kind of structural order (see below for details) that generates the exponential tail.

The difference in the structure related to the network formers and the modifiers is not only seen at short distances, but is
also reflected in the decay length of the structural correlation at large $r$ in that we find that the decay of the NaNa correlation is faster than the SiSi correlation, see values given in panels (f) and (h).
This result can be attributed to the fact that the Na atoms are only weakly attached to the Si-O network which gives them a larger structural flexibility, resulting in a smaller correlation length. Below we will discuss the temperature and composition dependence of this structural decay behavior in more detail.

To verify whether $|g_{\alpha\beta}(r)-1|$ decays at large distances indeed in an exponential manner, i.e., to test whether there are features in the correlation function that hint the presence of a non-trivial MRO, we have multiplied the curves for 300~K by an exponential to make the data more or less horizontal, see Figs.~\ref{fig_gr-1_nsx-l03-SiX}(i)-(l). One 
observes that the resulting curves show no noticeable signature for a secondary modulation of the partials, which indicates that from the point of view of $g(r)$ there is no signature of a non-trivial MRO.

\subsection{Revealing orientational structural order by four-point correlation functions}
The structural quantities that we have discussed so far have been widely used in the past since they allow to gain important insight into the structure of the silica and sodium silicate glass-formers at relatively short length scales. However, the rings are basically a topological quantity, while the static structure factor and the radial distribution functions seem not to reveal unexpected
information on the structure on larger distances. It is therefore of interest to investigate whether higher order correlation functions are able to uncover a non-trivial MRO and to provide insight into the structure of the glass in three dimensions. In the following, we hence present the results concerning the 3D structure of the liquids and glasses using the four-point correlation functions that have been introduced recently~\cite{zhang2020pnas}, see Eqs.~(\ref{eq7})-(\ref{eq10}), and which have demonstrated the ability to reveal in a variety of disordered systems new 3D structural features~\cite{zhang2020pnas,yuan2021packing,singh2023intermediate,tang2023friction}. 

\begin{figure}[ht]
\center
\includegraphics[width=0.68\textwidth]{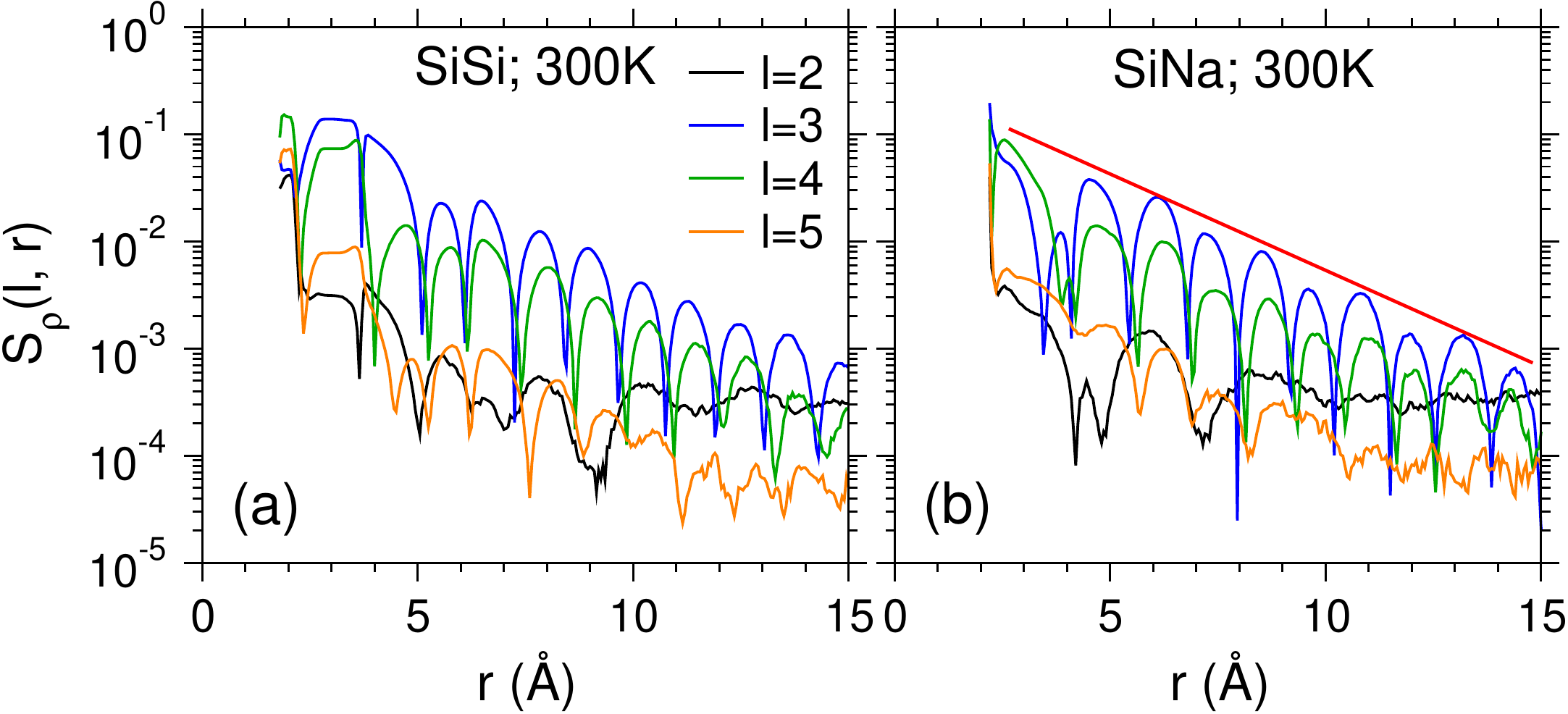}
\caption{$r$-dependence of $S_\rho(l, r)$ for NS5 at 300~K. (a) and (b) are, respectively, for the SiSi and SiNa correlations. The curves correspond to different values of $l$ and one sees that the mode $l=3$ is most pronounced, indicating a dominant tetrahedral order in the structure. The red solid line in panel (b) helps to see the secondary modulation (i.e., high/low peaks) of the $l=3$ curve.
}
    \label{fig_srho_n5-ldepend-SiX}
\end{figure}

\begin{figure*}[ht]
\center
\includegraphics[width=0.9\textwidth]{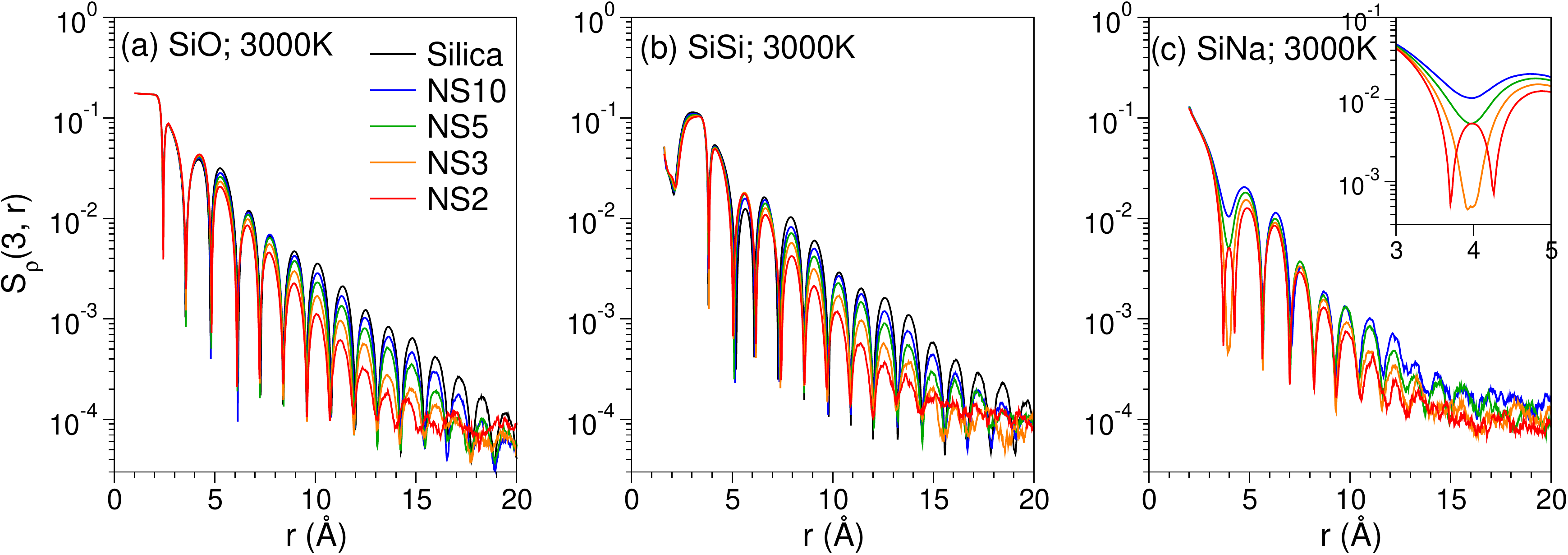}
\includegraphics[width=0.9\textwidth]{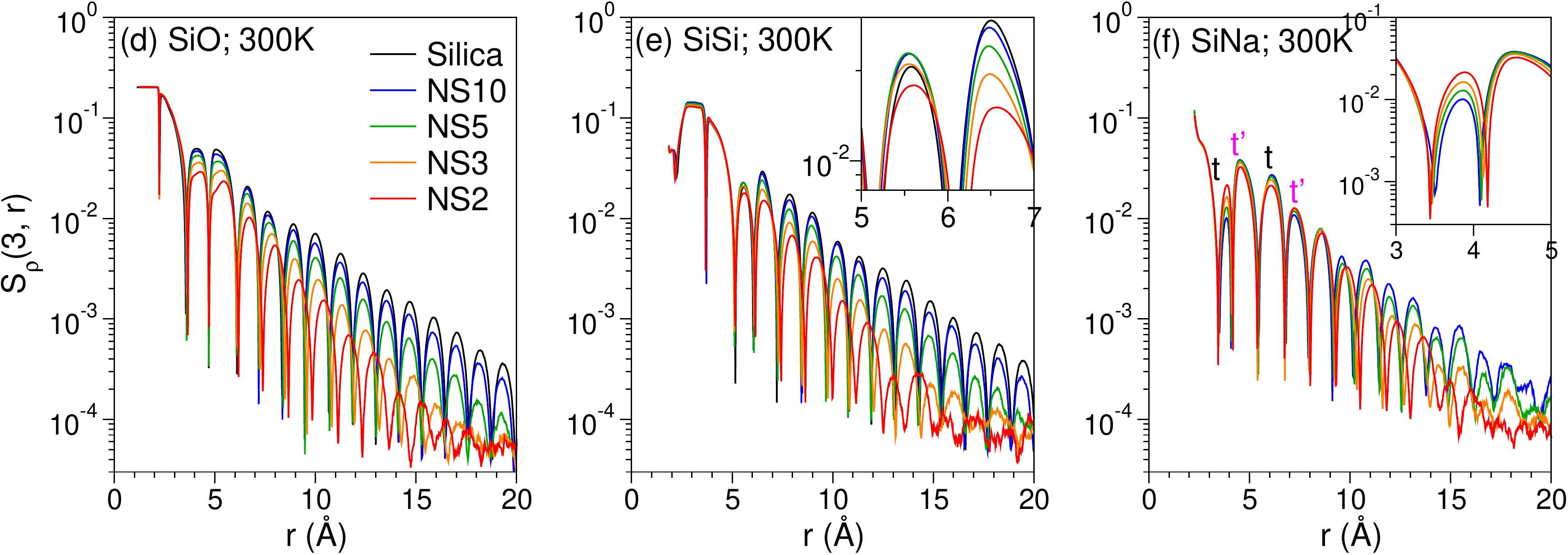}
\includegraphics[width=0.9\textwidth]{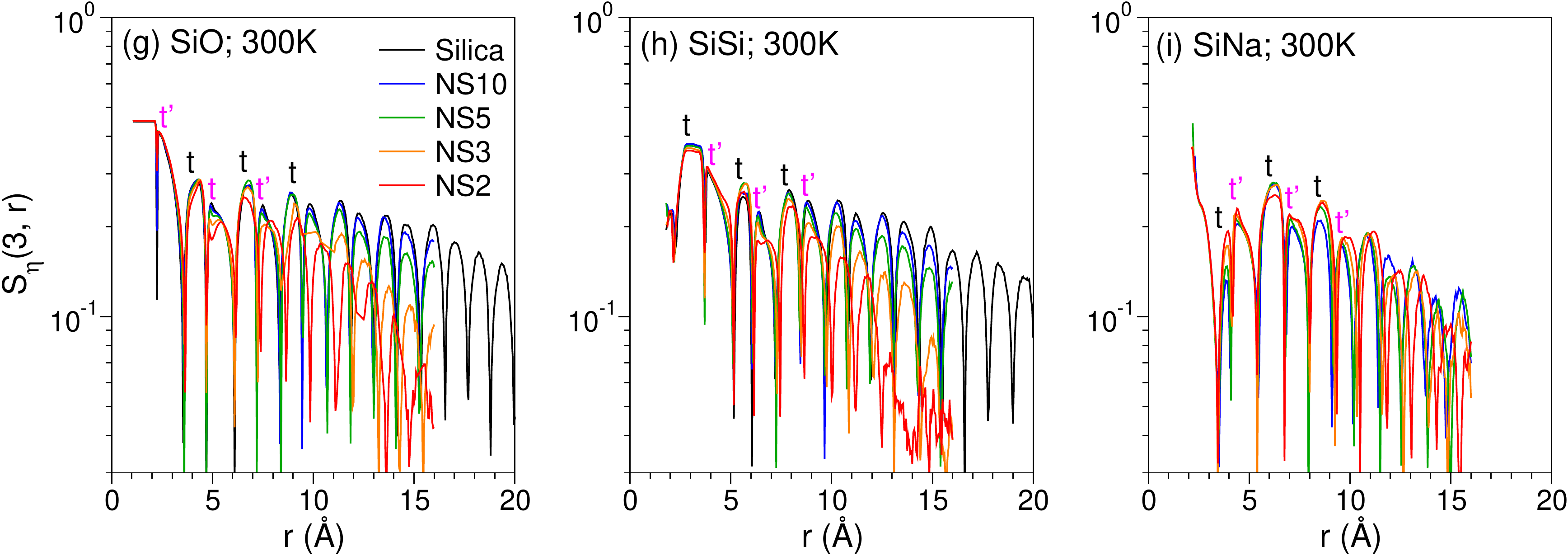}
\caption{Four-point correlation functions. Top panels: $S_\rho(3, r)$ for NSx  at 3000~K. The inset in (c) shows the enlarged view of a composition-induced transition behavior of $S_\rho(3, r)$ at $r\approx4$~\AA\ which is not seen at all in $g(r)$. Middle panels: The same quantities at 300~K. The inset in (e) shows a non-monotonic composition dependence of the orientational order at intermediate distances. Bottom panels: $S_\eta(3, r)$ at 300~K. The letters $t$ and $t'$ 
stand for tetrahedron and anti-tetrahedron, respectively. From left to right, the panels are for the SiO, SiSi, and SiNa correlations, respectively. \\ 
}
\label{fig_srho_seta_nsx-l03}
\end{figure*}

The quantity we probe is $S_{\rho}(l,r)$, i.e., the amplitude of the mode for index $l$ at distance $r$ of the four point correlation function, see Eq.~(\ref{eq9}). We recall that the index $l$ is sensitive to the nature of the symmetry of the density distribution on the sphere and thus gives insight on the 3D distribution of the particle field.
Figure~\ref{fig_srho_n5-ldepend-SiX} presents for different values of $l$ the $r$-dependence of $S_{\rho}(l,r)$ for the SiSi and SiNa correlations of the NS5 glass. (Thus a O-Si-O triplet is used to define the local coordinate system and one probes the 3D distribution of the Si and Na atoms in this coordinate system.) One recognizes that the mode $l=3$ has the strongest signal, a result which is reasonable since the system is dominated by local tetrahedral order and the $l=3$ mode is the best to capture this symmetry. (Note that this is in contrast to the case of hard-sphere like systems, such as metallic glasses and granular materials, in which one usually finds a dominating icosahedral symmetry which is best probed by the $l=6$ mode~\cite{zhang2020pnas}.) In the following we will hence focus on the $r$-dependence of this quantity.

Figure~\ref{fig_srho_n5-ldepend-SiX} shows that the $S_{\rho}(l,r)$ for the other modes are qualitatively very similar to the $l=3$ curves and hence this choice is not crucial, in other words, the slope of the envelope, and hence the decay length, is independent of $l$. However, we note that the peak positions do depend on $l$ since different $l$ capture different symmetries and the latter will depend on the distance $r$ from the central particle, see below. Finally we mention that the peak height of $S_\rho$ for SiNa shows a remarkable modulation in that it decays in a double step fashion. This feature is not visible in $g_{\alpha\beta}(r)$ and demonstrates that $S_\rho(l,r)$ allows to reveal structural properties in the MRO that are not accessible in two-point correlation functions. Below we will see that this modulation is related to a tetrahedral orientational order in the glass that alternates with  $r$.

Figures~\ref{fig_srho_seta_nsx-l03} shows the partials of $S_{\rho}(3,r)$ for the NS$x$ liquids at 3000~K [panels (a)-(c)] and at 300~K [panels (d)-(f)]. Overall, one observes that the $r$-dependence of SiO and SiSi correlations are very similar, which can be attributed to the fact that Si and O are strongly coupled to each other via a covalent bond. These curves show only a mild dependence on Na concentration (they become steeper with the addition of Na), indicating that the depolymerization of the network does not lead to a relevant change of the structural symmetry. Both functions display at intermediate and large distances a clear exponential decay, also seen in the SiNa signal [panels (c) and (f)], which allows to define a decay length $\xi$ that will be discussed in more detail below.
For the SiNa correlation at high $T$, one also notes an interesting feature at $r \approx 4$~\AA, in that the function changes its shape by splitting one minimum into two if the Na concentration is increased, see the Inset in panel (c). This distance corresponds approximately to the mid-point from the first peak to the first valley in $g_{\rm SiNa}(r)$ (see Fig.~\ref{fig_gr-1_nsx-l03-SiX}(c)), i.e., where $g_{\rm SiNa}(r)$ is close to 1.0. Therefore the described change in $S_\rho(l,r)$ indicates that at this $r$ the symmetry of the Na configuration changes qualitatively. 
No such change is observed in $g_{\rm SiNa}(r)$, see Fig.~\ref{fig_gr-1_nsx-l03-SiX}(c), demonstrating that $S_\rho$ is a powerful quantity which allows to unravel interesting structural information that is not accessible in the two-point correlation functions. We also note that at low $T$, Inset in Fig.~\ref{fig_srho_seta_nsx-l03}(f), no such transition in the symmetry is seen.

\begin{figure*}[th]
\center
\includegraphics[width=0.8\textwidth]{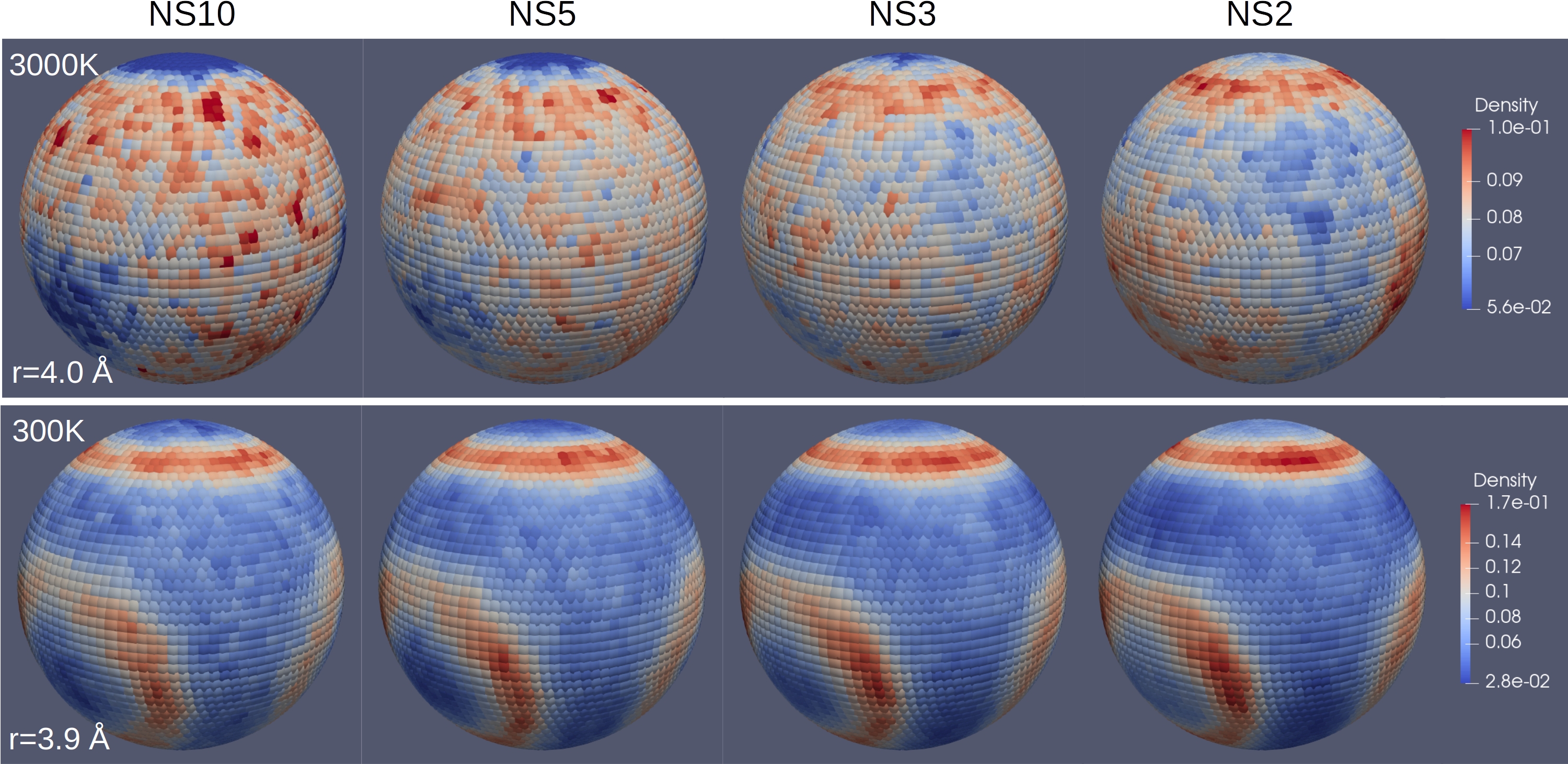}
\caption{Density distribution $\rho(\theta,\phi,r)$ for the SiNa correlation at $r \approx 4$~\AA, the distance at which one observes a transition in $S_\rho(3, r)$, Fig.~\ref{fig_srho_seta_nsx-l03}(c). Note that this is a short range structure. Top panels: $T=3000$~K. Bottom panels: $T=300$~K. 
}
\label{fig_densityplot-nsx-sina}
\end{figure*}

To elucidate the nature of the discussed symmetry change at $r\approx 4$~\AA, we probe directly the particle density distribution in 3D. This distribution is presented in Fig.~\ref{fig_densityplot-nsx-sina}, and confirms that at high $T$ (upper panels) the symmetry of the density field changes with increasing Na concentration: For low concentration one finds large areas with significant Na density (reddish colors) while for NS2 these regions become depleted (turning into blue) and instead one notices 4 rings that have higher particle density and that form a tetrahedral structure, thus explaining the growth of the peak in $S_\rho(3,r)$ at this distance. 
We conjecture that this ordering transition is due to the repulsion between the Na atoms: If the Na concentration is low the Na atoms will occupy a large area (red zone for NS10) since this is entropically advantageous and energetically (repulsive Na-Na interactions) not strongly penalized since at this $r$ the number of Na atoms is relatively small and hence they can avoid each other. With increasing Na concentration the repulsion between the Na atoms locks them into a tetrahedral symmetry, thus leading to the increased particle density in the mentioned (red) rings in NS2, and at the same time to the depletion of the (red) zones that were occupied in the NS10 system. In other words, the energetic term leads to an ordering of the Na configuration at the expense of a lower entropy.  
At low $T$, see lower panels of Fig.~\ref{fig_densityplot-nsx-sina}, the entropic contribution to the free energy is significantly decreased and as a consequence the energetic term always dominates, resulting that the arrangement of the Na atoms is given by the mentioned four rings, and this independent of the Na concentration. This agrees with the data presented in the Inset of Fig.~\ref{fig_srho_seta_nsx-l03}(f) which confirms the absence of a structural transition.
Finally, we emphasize that the symmetry change at this distance cannot be simply ascribed to the density difference between the melts,  
see Fig.~\ref{fig_rho_Tg}, but is related to a change in the underlying local structure which the standard structural observable are not able to detect. Below we will see that this structural transition can also be induced by the change of temperature.

The change in the SiNa correlation is not the only 
structural modification that can be detected with the help of the four point correlation functions. Figure~\ref{fig_srho_seta_nsx-l03}(e) shows that the height of the peak in the SiSi correlation at $r\approx 5.5$~\AA\; changes in a non-monotonic manner as a function of the Na concentration (see Inset), indicating a complex composition dependence of the MRO which is not detected in $g_{\rm SiSi}(r)$, see Fig.~\ref{fig_gr-1_nsx-l03-SiX}(f). 
A closer inspection of the structure in real space reveals that this composition dependence is due to a non-monotonic change of the strength of the tetrahedral orientational order at this distance with increasing Na concentration (the maximum is seen for the composition between NS10 and NS5). 

As already mentioned in Sec.~\ref{sec:methods}, the decay of $S_\rho(l,r)$ with increasing $r$ is not only due to the loss of symmetry, but also to a decreasing amplitude of the density fluctuations. To disentangle these two contributions it is therefore useful to consider the 
four-point correlation function $S_\eta(3, r)$, i.e., the normalized version of $S_\rho(3, r)$, see Eq.~(\ref{eq10}). For the glasses at 300~K, Figs.~\ref{fig_srho_seta_nsx-l03}(g)-(i), $S_\eta(3, r)$ shows peaks with a height that alternates, with the high/low peaks corresponding to the tetrahedral (labelled by $t$) and anti-tetrahedral ($t'$) symmetries, respectively. (Anti-tetrahedral means here the tetrahedron that is dual to the central tetrahedron.) This modulation of peak heights originates from the fact that these two kind of peaks do have the same symmetry, but not the same distribution of the particle density, i.e., the tetrahedral peak has a better defined symmetry than the anti-tetrahedral peak. The effect of the alternating height is more visible in $S_\eta$ than in $S_\rho$ since the former considers the normalized densities, i.e., it is more sensitive regarding the shape of the density distribution instead of its amplitude. For distances larger than $r \approx 12$~\AA, the modulation is no longer seen in $S_\eta$, indicating that this is a MRO effect. Below we will see that this alternating sequence gradually disappears with increasing $T$.

\begin{figure*}[ht]
\center
\includegraphics[width=0.9\textwidth]{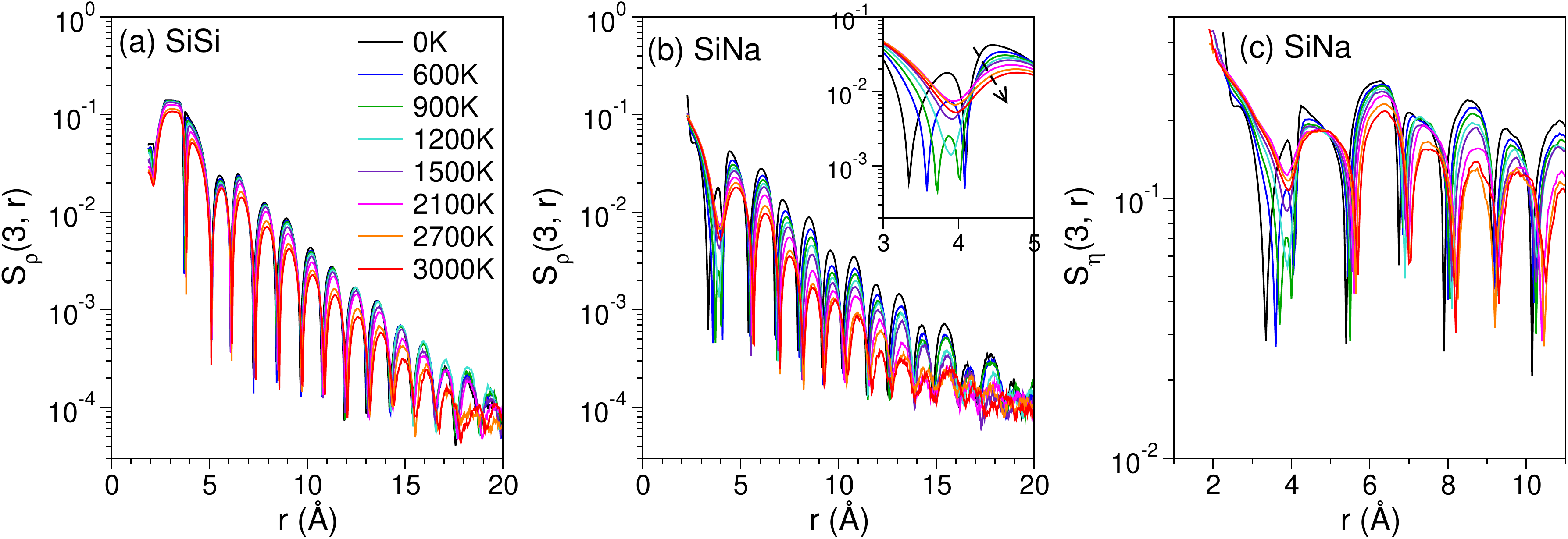}
\includegraphics[width=0.9\textwidth]{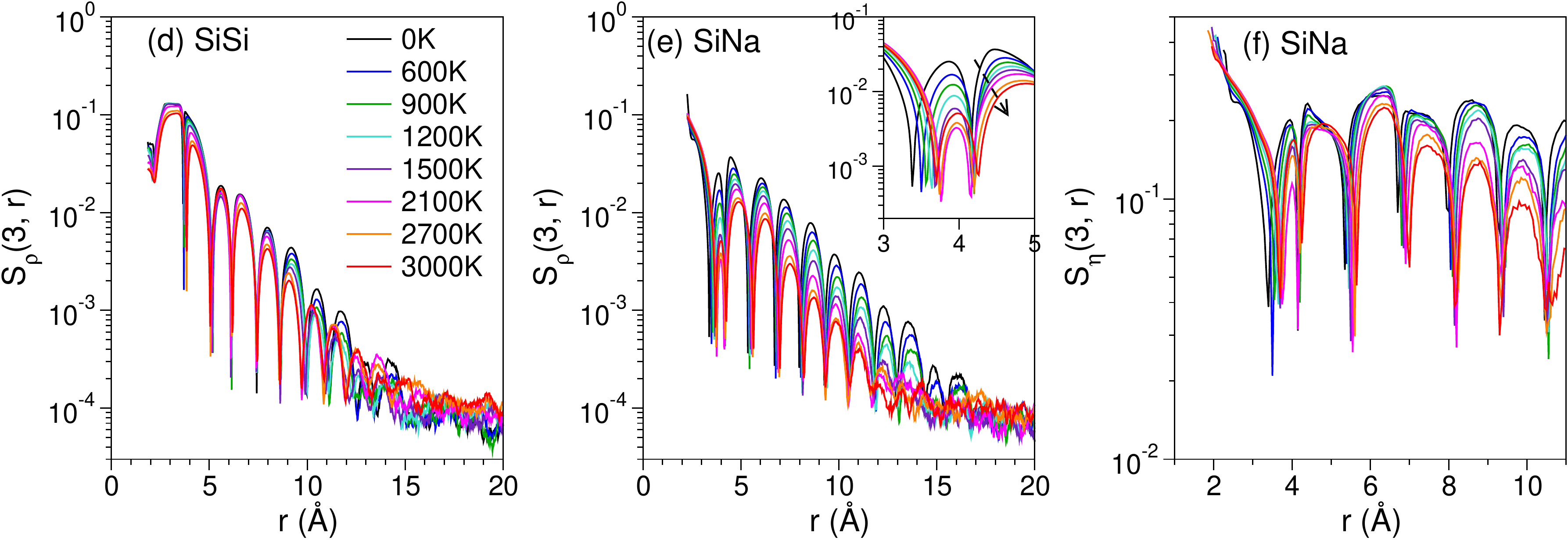}
\caption{$S_\rho(3, r)$ and $S_\eta(3,r)$ at different temperatures for NS5, (a)-(c), and NS2 (d)-(f). The Inset in (b) is a zoom  of $S_\rho(3, r)$ at $r\approx4$~\AA\, showing the transition in structure at around $T=900$~K. No such transition is observed in NS2 (Inset of panel (e)).
$T$-dependence of $S_\eta(3,r)$ for the SiNa correlation. In panels (c) and (f), the alternating height of the peaks disappear as $T$ increases. 
}
    \label{fig_srho_l03-SiX-Tdepend}
\end{figure*}

\begin{figure*}[ht]
\center
\includegraphics[width=0.9\textwidth]{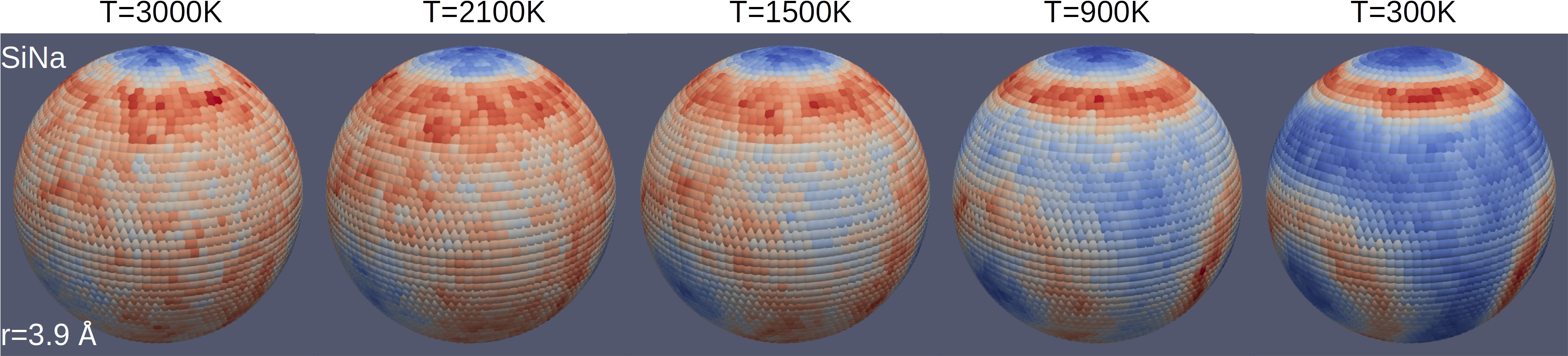}
\caption{Temperature dependence of the density distribution $\rho(\theta,\phi,r)$ for the SiNa correlation at $r=3.9$~\AA\ for the NS5 glass.
Decreasing temperature leads to a change of symmetry corresponding to the transition seen in $S_\rho(3, r)$, Inset of Fig.~\ref{fig_srho_l03-SiX-Tdepend}(b). Red and blue correspond to high and low particle densities, respectively. The range of the color map is adapted for each frame.
     }
\label{fig_rho-ns5-sina-Tdepend}
\end{figure*}

Having discussed the nature of the MRO as a function of composition, we now focus on its temperature dependence. Figure~\ref{fig_srho_l03-SiX-Tdepend} shows the $T$-dependence of $S_\rho(3, r)$ for NS5 [panels (a)-(c)] and NS2 [panels (d)-(e)], respectively. One recognizes that the curves for SiSi show a smooth and mild evolution with $T$, i.e., no new features emerge if $T$ is decreased, and the same behavior is found for the SiO correlation (not shown). This is in contrast to the SiNa correlation which shows a marked $T$ dependence and, for the case of NS5, panel (b), the formation of a new peak at $r \approx 4$~\AA, see Inset. The temperature at which this peak starts to appear is around $T=900$~K and with decreasing $T$ its amplitude increases substantially.  Note that this change of the $r-$dependence is also seen for $l=4$ (not shown), although less pronounced, indicating a significant modification of the local structure. We also mention that no such change is found in NS2, see Inset in panel~(e), i.e., this effect is related to the relatively low concentration of the Na atoms. A real space representation of this temperature-induced transition behavior is shown in Fig.~\ref{fig_rho-ns5-sina-Tdepend}. The change of symmetry with temperature is smooth but clearly visible. The mechanism leading to this transition is the same as the one discussed in the context of Fig.~\ref{fig_densityplot-nsx-sina}, namely the competition between an entropic term that favours a more uniform distribution of the Na atoms and an energetic term that induces a clustering of these atoms. Hence, we conclude that the transition from the delocalized positions of the Na atoms can either be induced by a change of temperature or by a change of Na concentration. 

To probe how the change of the local structure affects the MRO, we present in Fig.~\ref{fig_srho_l03-SiX-Tdepend} the $T$-dependence of $S_\eta(3,r)$ for NS5 [panel (c)] and NS2 [panel (f)], respectively. One confirms that the transition at $r \approx 4$~\AA\ occurs at around 900~K, i.e., deeply in the glass phase ($T_{\rm g} \approx 1950$~K), which indicates that the transition is of vibrational/elastic nature. Furthermore, one notes that also the secondary modulation of the MRO starts to become pronounced at around 900~K. Hence, it is likely that the enhanced ordering of Na at short-range distances (as seen from the symmetry change at $r \approx 4$~\AA) triggers the ordering of the structure at larger distances.

\subsection{Structural corelation lengths}

As mentioned above, we find for distances beyond the short-range that the envelope of $|g(r)-1|$ as well as $S_\rho(r)$ exhibit an exponential decay, in agreement with findings for other systems~\cite{zhang2020pnas}. However, we also note that the decay of $g(r)$ is often described by the functional form

\begin{equation} 
\label{eq_exp-decay-gr}
    |g(r)-1| \propto \frac{{\rm exp}(-r/\xi_g)}{r} \quad,
\end{equation}

\noindent
which is motivated by the Ornstein and Zernike relation~\cite{ornstein1914influence}. 
Here, $\xi_g$ is the structural correlation length which can be obtained by fitting the peaks of these correlation functions using this functional form. Similarly the decay of $S_\rho(3,r)$ can be described by

\begin{equation} 
\label{eq_exp-decay-srho}
    S_\rho(3,r) \propto \frac{{\rm exp}(-r/\xi_\rho)}{r} \quad,
\end{equation}

\noindent
where $\xi_\rho$ is the corresponding structural correlation length. We find that Eqs.~(\ref{eq_exp-decay-gr}) and~(\ref{eq_exp-decay-srho}) describe very well 
the decay of the correlation functions at intermediate and large distance, thus allowing to determine the structural correlation lengths $\xi_g$ and $\xi_\rho$.  For $S_\rho(3, r)$, we have chosen for the lower limit of the fitting range $r=5$~\AA\ for the Si-O pair and $r=6$~\AA\ for the Si-Si and Si-Na pairs, while the upper limit was determined by considering only $S_\rho> 3 \times 10^{-4}$ (to exclude data with strong noise). For $|g(r)-1|$ the fit was done in the range $r>7$~\AA\ and $|g(r)-1|>0.004$. We note that the so-obtained  correlation lengths are stable with respect to a moderate change of the fitting ranges.

\begin{figure*}[ht]
\center
\includegraphics[width=0.9\textwidth]{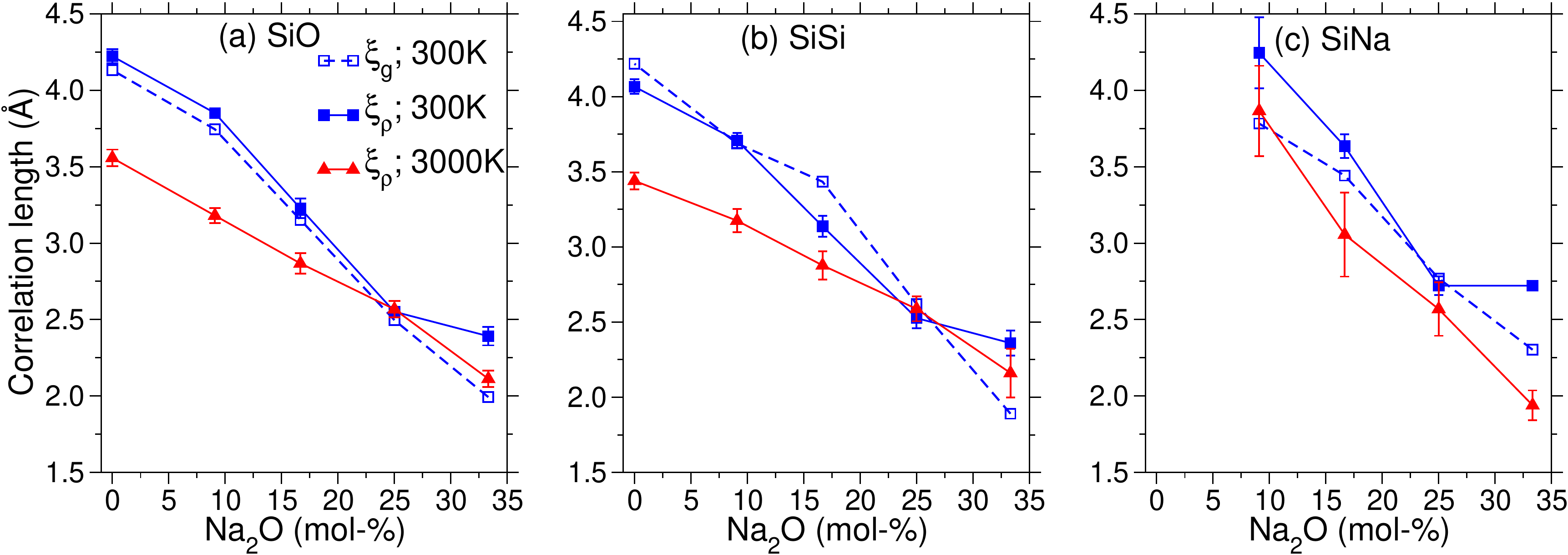}
\includegraphics[width=0.9\textwidth]{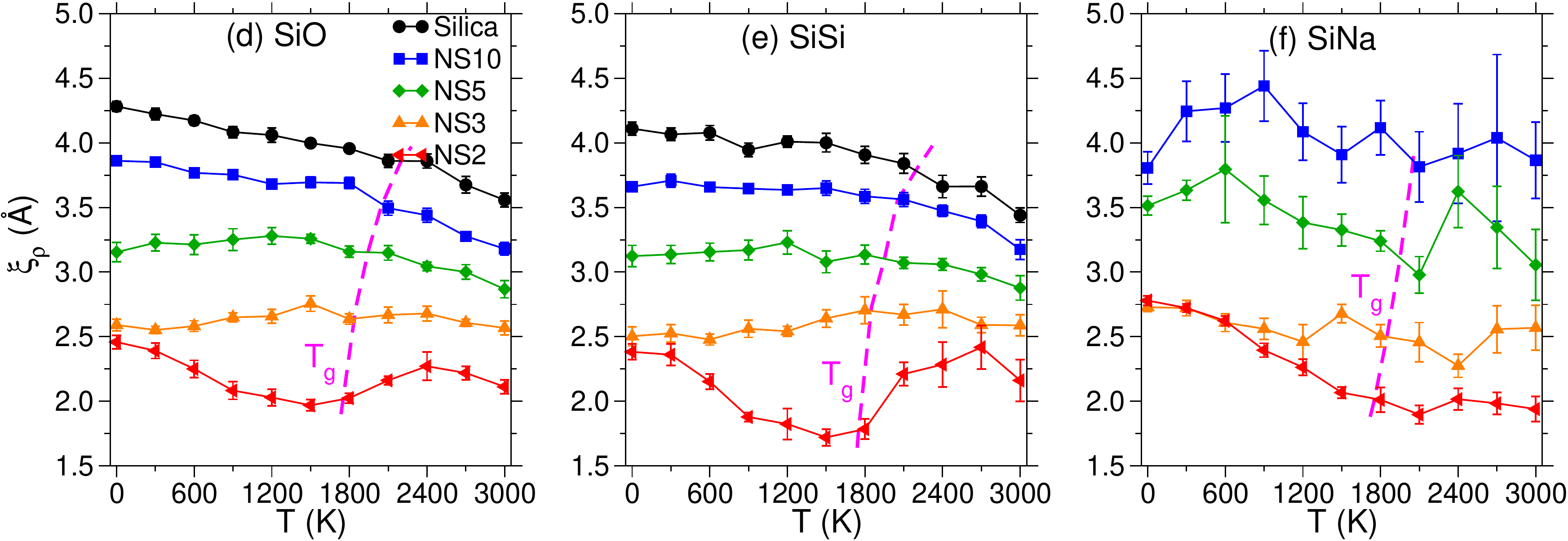}
\caption{Decay length $\xi$ as a function of composition, (a)-(c), and temperature, (d)-(f). 
Error bars represent the standard error of the mean of 8 independent samples. The approximate glass transition temperature $T_{\rm g}$ as estimated from the simulations is indicated by the pink dashed line.
}
    \label{fig_decay-length-nsx}
\end{figure*}

Figure~\ref{fig_decay-length-nsx} presents for the various correlations the composition-and temperature-dependence of this decay length. We note that $\xi_g$ and $\xi_\rho$ as extracted from $g(r)$ and $S_\rho(3, r)$ are quantitatively compatible with each other, see panels (a)-(c), consistent with the finding of previous studies~\cite{zhang2020pnas,yuan2021packing}. This result can be rationalized by recalling that $g(r)$ is closely related to the angular average of the four-point correlation functions and thus they have similar decay behavior. However, the four-point correlation function has the advantage that its signal to noise ratio is significantly better than the one of $|g(r)-1|$, thus permitting to obtain reliable structural data up to larger distances. Furthermore $S_\rho$ allows to study the orientation-dependence of the particle density, i.e., a novel structural information. 

Figures~\ref{fig_decay-length-nsx}(a)-(c) also demonstrate that these scales depend strongly on composition in that they change by about a factor of two if the Na concentration is increased from zero to 33\%. The correlation lengths for SiO and SiSi, panels (a) and (b), are basically identical, a result that is reasonable since Si and O constitute the highly correlated network. The decay length of the SiNa correlation, panel (c), is similar to the two other scales, although its dependence on Na concentration is a bit more pronounced, in particular at high $T$. This latter observation can be related to the high flexibility of the modifiers to adapt to the disordered network structure. 

For the liquids at 3000~K, one finds a surprisingly linear $f$-dependence of $\xi$. At small Na concentration, such a dependence is expected since the addition of Na will cut Si-O bonds which allows the silica matrix to restructure and as long as these cuts can be considered as independent processes, the correlation length will decrease linearly with the number of defects, i.e., cut bonds. That this linear decay is observed even at large Na concentration indicates that at high temperatures the depolymerization of the network structure can be considered to be close to a random process, i.e., the interaction between the Na atoms is not very relevant. This conclusion is in line with the results presented in Figs.~\ref{fig_cn-nsx-sina-nana}(a) and (c) where we showed that the distribution of the Na-coordination numbers does indeed follow closely a random distribution which indicates that the Si-Na coordination number between nearby Si atoms are at most weakly correlated.

For the glasses at 300~K, the $f-$dependence of $\xi$ is still roughly linear, in agreement with the argument presented above for high temperatures. 
One notes, however, that the $\xi$ for Si-O and Si-Si show a clear bend at around 10\% of Na$_2$O. 
This bend signals that beyond this Na concentration  
the domains in the Si-O matrix that are not significantly affected by the Na atoms start to shrink rapidly with $f$ and hence the correlation length decays faster, indicating the presence of a percolation-like phenomenon of the perturbation.
The existence of the bend is also coherent with the cross-over in the composition at which one observes a transition in the non-linear elasticity of the NSx glasses~\cite{zhang2022stiffness}. This concurrence hints that at this composition the defect-free domains start to become mechanically decoupled from their neighboring domains, causing a change of the elastic properties of the glass. In the following section we will see that the decay length is indeed correlated with the (macroscopic) mechanical properties of the system, thus bolstering this conclusion. Since at the bend the value of $\xi$ is around 3.5~\AA\;, one can conclude that at the decoupling the size of these domains is on the order of 10-12~\AA\ (assuming that this size is given by two to three times $\xi$, i.e., when the structural correlation has basically vanished).

Figures~\ref{fig_decay-length-nsx}(d)-(f) presents the $T$-dependence of the different $\xi$'s, and we start our discussion by focusing on the length scale for the SiO and SiSi correlations since they characterize the properties of the network structure. For the systems with low Na concentration, $\xi$ increases with decreasing temperature, a result that is reasonable since upon cooling the liquid becomes more structured. For silica and NS10 we can distinguish two regimes: At high $T$ the growth of $\xi$ with decreasing temperature is relatively fast while at low $T$ it is milder. The cross-over between these two regimes is close to the (simulation) $T_g$, marked by a magenta dashed line. This indicates that the high $T$ regime is related to a topological change in the structure, i.e., the way particles are packed, while the $T < T_g$ regime is due to elastic relaxation.

For intermediate concentrations of Na, NS5 and NS3, the increase of $\xi$ at high temperatures is still observed, but below $T_g$ the length scale becomes basically flat or shows even a weak tendency to decrease. A hint  regarding the origin of this behavior comes from the curve for NS2 which shows an even more complex $T$-dependence: At high $T$ the length scale increases as expected. However, at around 2400~K, $\xi$ shows a maximum and decreases by about 20\% (for SiSi) if $T$ is reduced to 1500~K before it increases again at even lower temperatures. Note that the temperature at which one observes the local maximum is well above $T_g$, i.e., it is not an out-of-equilibrium effect but instead reflects a change in the equilibrium structure of the system. This $T$-dependence can be rationalized as follows: At high $T$ the system is quite homogeneous and upon lowering the temperature it will behave like any other liquid and hence become more structured if $T$ is lowered~\cite{zhang2020pnas}, i.e., $\xi$ will increase. However, the presence of the alkali atoms perturbs the ideal random network structure of the Si-O matrix, i.e., increases its enthalpy, which induces a thermodynamic driving force that generates zones in the sample in which the concentration of Na atoms is lower/higher than the nominal composition, resulting in a reduction of the enthalpy. 
Although with respect to the entropy this process is not favorable, the gain in enthalpy due to the creation of Si-O domains with fewer defects is sufficiently large to permit the system to lower its free energy. This mechanism makes that the sample becomes structurally more heterogeneous with decreasing $T$, thus explaining why the length scale $\xi$ decreases. (We emphasize that this process should not be confused with an arrested demixing, but rather be seen as a modulation of the local composition that occurs in equilibrium.) Although at the (simulation) $T_g$, the silica matrix becomes rigid, the Na atoms can still move to some extent because they are only weakly coupled to the matrix, i.e., the mentioned process is not completely halted but only slowed down and hence $\xi$ will continue to decrease even below $T_g$. At temperatures significantly below $T_g$ also the motion of the Na atoms becomes frozen and only elastic relaxation can occur, resulting in the increase of the correlation length with decreasing temperature, in agreement with the NS2 curve at low $T$.

It can be expected that the temperature at which the balance between enthalpic and entropic forces is reached, i.e., where $\xi$ shows a maximum, will depend on the composition. Decreasing the concentration of Na will make that the thermodynamic driving force for generating the heterogeneous structures is weakened which makes that the transition will occur at lower temperatures. This rationalizes thus the observation that the maximum in $\xi$ shifts to lower temperatures with deceasing Na concentration, in agreement with the trend seen in the $\xi$ data for NS2, NS3, and NS5. Since a decrease in the Na concentration makes that $T_g$ rises, the local minimum in $\xi$ seen in NS2 becomes washed out, since it occurs at a temperature that is below $T_g$. Finally, we mention that the mechanism giving rise to the non-monotonic $T$-dependence of $\xi$ is indeed compatible with the concentration-dependence of $\xi$ at $T=300$~K, as discussed in the context of panels (a) and (b).

Panel (f) of Fig.~\ref{fig_decay-length-nsx} shows the decay length for the SiNa correlation. Due to the low concentration of Na the $\xi$ for NS10 and NS5 is rather noisy and is basically independent of $T$. In contrast to this, one can identify for NS3 and NS2 a clear $T$-dependence: At high temperatures $\xi$ is constant, with a weak minimum at 2400~K (NS3) and 2100~K (NS2), while at lower $T$ one observes an increasing $\xi$. For the case of NS2, the temperature at which the $T$-dependence changes is compatible with the temperature at which also the length scales for SiO and SiSi start to increase. This is thus evidence that at $T$'s below this temperature, the growth of $\xi$ is due to the local ordering of the glass structure via elastic processes. In contrast to this, the weak $T-$dependence of $\xi$ at higher temperature indicates that the arrangements of the Na atoms does not change significantly with $T$. The $T$-dependence of $\xi$ for NS3 is compatible with this view, although the effect is less pronounced.

We also note that the non-monotonic $T$-dependence of $\xi$ is not due to the fact that the density of the system changes with temperature (see Fig.~\ref{fig_rho_Tg}). We have calculated the scaled structural correlation length $\xi'(T)=\xi(T)[\rho(T)/\rho(3000{\rm K})]^{(1/3)}$ (not shown) to take into account the effect of thermal expansion and found that $\xi'$ shows a similar $T-$dependence as $\xi$, although this dependence becomes stronger with increasing Na concentration; the largest difference between $\xi'$ and $\xi$ is found to be around 15\% for the NS2 glass at zero temperature. Hence we conclude that the non-trivial $T-$dependence of $\xi$ reflects a structural rearrangement of the atoms that is not detectable from the density.

Our finding that $\xi$ is non-monotonic in temperature indicates that the proposal by Ryu {\it et al.} that  the correlation length characterizing the MRO can be used to predict the properties of the ideal glass state~\cite{ryu2019curie} is not working for soda-silicate glasses, and most likely for other alkali-silicates neither. This is due to the fact that in these glass-formers the structural properties are influenced by several competing mechanisms, thus making an extrapolation of the structure to low temperatures very difficult. (Ryu {\it et al.} considered only metallic glass-formers for which there seems to be only one relevant mechanism since their structure is relatively simple.) This unexpected non-monotonic dependence of the MRO is also at variance with the starting hypothesis of several theoretical frameworks, such as the mode-coupling theory or the random first-order transition theory, 
that describe the slowing down of the dynamics of glass-forming systems~\cite{binder_kob_2011,cavagna2009supercooled}. Most of these theories are based on the behavior of simple liquids, i.e., hard-spheres, Lennard-Jones, etc., that have a simple dependence of their structure as a function of temperature. How a non-monotonic dependence of $\xi$ affects the slowing down of the dynamics is therefore an open question. It is clear, however, that the influence of the non-monotonic behavior of $\xi$ on quantities like the viscosity must be small, since for oxide glasses the latter is known to increase continuously with decreasing $T$~\cite{mazurin_handbook_1983}. Nevertheless, it is conceivable that the $T-$dependence of the activation energy reflects the subtle modulation of the correlation length. Furthermore, it is reasonable to assume that the conductivity of ion-conducting glasses and melts will depend on the geometry of the conducting pathways, structures that are certainly related to the MRO. It will thus be interesting to probe in the future these quantities in order to see whether they reveal information on the MRO.

\subsection{Correlating the MRO with macroscopic properties of liquids and glasses}

Having presented our findings regarding the MRO in sodo-silicate glasses, we now address the question how this order is correlated with the macroscopic properties of liquids and glasses. Earlier studies have given evidence that the MRO structure correlates well with one of the most important parameters describing the dynamics of  glass-forming liquids, namely the kinetic fragility~\cite{mauro_structural_2014,
ryu2020origin,yu2022silica}, and also with the elastic properties of the glasses as characterized by the Poisson ratio, $\nu$~\cite{rouxel_elastic_2007,greaves_poissons_2011}. 
 
To establish these correlations, various observables have been used to quantify the MRO, with some of them being rather indirect, such as the ring size distribution~\cite{shi2023revealing} or the dimensionality of the network~\cite{rouxel_elastic_2007}.
It is thus important to probe whether the characteristic decay length as obtained from $S_{\rho}$ (Fig.~\ref{fig_decay-length-nsx}) correlates with these observables. 

\begin{figure}[ht]
\center
\includegraphics[width=0.68\textwidth]{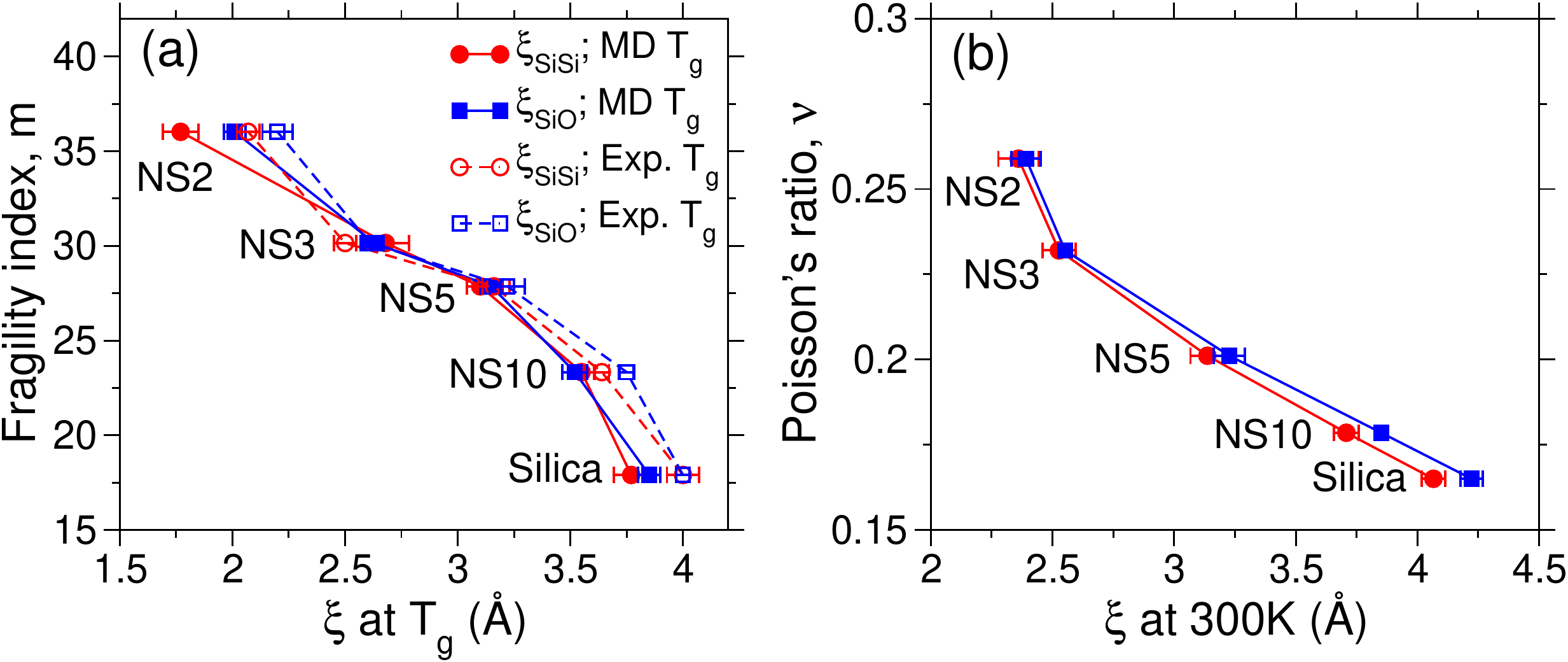}
\caption{Correlating the structural decay length with the kinetic fragility of the glass-forming liquids, panel (a), and the Poisson's ratio of the glasses, (b). The data points are labeled with their corresponding composition. Experimental data for the fragility index, $m$, and the Poisson'ratio, $\nu$, of the glasses are taken from Refs.~\cite{nascimento2007viscosity,sidebottom2019connecting} and~\cite{bansal_handbook_1986}, respectively. 
}
    \label{fig_decay-length-fragility-poisson}
\end{figure}

The kinetic fragility is a property of the liquid in its equilibrium state at the glass transition temperature $T_g$ and is usually quantified by the fragility index $m=d({\rm log}_{10}\eta(T))/d(T_{\rm g}/T)|_{T=T_{\rm g}}$, where $\eta$ is the shear viscosity~\cite{nascimento2007viscosity}. 
Since our simulations do not allow to probe the equilibrium dynamics of the liquid at the experimental $T_g$, defined via $\eta(T_g)= 10^{12}$Pa~s, we use the experimental values of $m$. (Note that for NS10 and NS5 experimental measurements of $m$ are not available and we have obtained theses values by a linear interpolation of the data between NS4 and silica). Figure~\ref{fig_decay-length-fragility-poisson}(a) presents the fragility index $m$ as a function of the decay length at the simulation $T_{\rm g}$ (curves with filled symbols). The data for the Si-O and Si-Si correlations show that the two curves are indeed very similar, as expected from our discussion of Fig.~\ref{fig_decay-length-nsx}.  
Interestingly, one recognizes that $m$ depends almost linearly on $\xi$ for the compositions with high Na concentrations, whereas there is a notable downward bending of the curves at low Na concentrations. The cross-over composition (occurring between NS5 and NS10) is compatible with the critical composition at which one observes a transition for the NS$x$ glasses from a strain hardening behavior to strain softening~\cite{zhang2022stiffness}.  
These findings hint that the deformation properties of a glass are, at least partially, reflected in the kinetic fragility of the glass-forming liquid. Whether this connection is due to the fact that both quantities are correlated with the length scale characterizing MRO or whether there is another reason, remains an open question that should be studied in the future.

Many studies have shown that the relaxation dynamics of glass-forming liquids becomes increasingly cooperative, i.e., spatially heterogeneous, if the temperature is lowered~\cite{berthier_theoretical_2011,royall_role_2015}. How the size of the domains of these cooperatively rearranging regions depend on temperature or its relation with kinetic fragility are still open questions. 
Since recent experiments hint that the relaxation dynamics is significantly affected by the MRO~\cite{singh2023intermediate}, it is reasonable to assume that $\xi$ is related to the dynamical properties of the system. Following this logic one can thus conclude from Fig.~\ref{fig_decay-length-fragility-poisson}(a) that more fragile systems have a cooperativity length scale that is smaller than the one of strong glass-formers. Although this results seem to be at odds with the view that, e.g., in silica the relaxation dynamics is closely related to bond switching, i.e., a very local process, such a simple view neglects the importance of elastic effects. Since for highly polymerized networks these effects are evidently relevant on larger length scales, it is reasonable to conclude that the relaxation dynamics, and hence the fragility, is not only determined by the local motion of the particles, but also by structural features on larger length scales, in other words that the effective cooperativity length scale is large, in agreement with the findings of Fig.~\ref{fig_decay-length-fragility-poisson}(a).
Finally, we mention that the $\xi$ we use for this graph is obtained at the simulated glass transition temperature, and hence it is important to test whether the obtained correlation is also present if one uses the value of $\xi$ at the experimental $T_g$. Therefore we have used the data from Fig.~\ref{fig_decay-length-nsx} to estimate $\xi$ at this temperature and present the resulting curve of $m(\xi(T_g))$ in Fig.~\ref{fig_decay-length-fragility-poisson}(a) as well. One sees that the data points still suggest a strong correlation between $\xi$ and the kinetic fragility, i.e., that this is a robust result.

The Poisson ratio $\nu$ is widely recognized as an important material parameter for technology and structural design. Hence we examine here  whether this property is correlated with the structural decay length of the glass, see Fig.~\ref{fig_decay-length-fragility-poisson}(b). 
One observes that $\nu$ exhibits a very clear monotonic decrease with increasing $\xi$, indicating that these two quantities are indeed correlated. The rationale for this finding is that due to their small correlation length the systems with high Na concentration can be seen as being composed of small structural domains and hence are, from the point of view of elasticity, not that different from an incompressible fluid, which has $\nu=0.5$. In contrast to this, systems with large structural correlation length have an elastic response that is very different from such a fluid, i.e., the $\nu$ is significantly smaller. Finally, we mention that although $m$ as well as $\nu$ decrease with $\xi$, their dependence on $\xi$ is different in that the bending of curves at around NS10 is only seen in $m$, whereas $\nu$ shows at this composition basically a linear dependence on $\xi$, thus showing that the elastic response is related to the MRO in a simple manner.

\section{Conclusion and Outlook}

In the present work we have used molecular dynamics simulations to investigate how the structure of two prototypical oxide glass-formers, namely silica and sodo-silicates, depends on composition and temperature. In particular, we focused on the evolution of the order present on intermediate length scales by using standard two-point correlation functions such the static structure factor, ring structures in the network, as well as a recently proposed four-point correlation function that is able to capture three-dimensional structural order.

Firstly, at short distances, we found that the local environment of Na becomes more ordered with increasing Na concentration, as can be concluded from the decreasing width of the distribution of the nearest neighbor number, once it is normalized by its mean. Secondly, at intermediate length scales, the static structure factor indicates the presence of a transition in the spatial distribution of Na from ``pockets'' to ``channels'' as Na concentration increases, signaling a growing inhomogeneity in Na distribution, which is also observed from the snapshots of the atomic configurations. These findings generalize thus the qualitative picture proposed long time ago by Greaves who speculated that alkali-silicates glasses have a channel structure in which the concentration of alkali (e.g., Na and K) atoms is enhanced~\cite{greaves1985exafs}.  

A further consequence of increasing Na concentration is that the Si-O network becomes progressively depolymerized, resulting in the broadening of the ring size distribution, i.e., a change of structure on intermediate length scales. The radius of gyration of these rings was found to be well described by a power-law with an exponent about 0.75 (depending only weakly on composition and temperature), indicating that the rings are more and more crumbled with increasing size. Obtaining a theoretical understanding of this value and whether it is universal for all network-glassformers is an open problem which should be addressed in the future. 

The four-point correlation function $S_{\rho}(3,r)$ reveals that at $r \approx 4$~\AA\; the local symmetry of the Na arrangement depends on temperature and composition, a feature that is not detectable in $g(r)$. This change in the local structure can be rationalized by the competition between entropic and energetic terms. In addition, $S_{\rho}(3,r)$ allows to detect that the glass sample shows on intermediate length scales (4-15~\AA) a modulation of the structure which is related to the change of local orientation of the tetrahedra and not visible in the standard two-point correlation functions.

A deeper understanding of the MRO is achieved by probing the decay of $g(r)$ and $S_\rho(3,r)$ at intermediate length scales. The structural correlation length $\xi$ extracted from these two correlation functions are compatible with each other, but the one extracted from $S_\rho(3,r)$ can be obtained with significantly higher accuracy, thus permitting to study how the MRO depends on $T$ and composition. While $\xi$ grows monotonically with decreasing $T$ if the Na concentration is low, one finds for the Na-rich NS2 systems a non-monotonic $T-$dependence of $\xi$. The minimum in $\xi$ for the correlations among the network atoms (i.e. SiSi and SiO) at around 1500~K can be attributed to the competition between enthalpy and entropy. These competing forces make that the sample forms at intermediate temperatures domains that have high/low concentration of Na, thus a small correlation length, and only at the lowest temperatures these regions start to order, i.e., $\xi$ increases.

The fact that the $\xi$ and Na coordination shows both a non-monotonic $T-$dependence indicates that SRO and MRO are intimately related to each other. However, the MRO can develop features of its own, such as the orientational order discussed above,  
which in turn can be expected to influence the mechanical and vibrational features of the system. Evidence for such an influence is that the structural correlation lengths are found to correlate well with macroscopic properties of the glass-formers such as the kinetic fragility or the Poisson'ratio. 

Apart from obtaining a deeper insight into the structural properties of glass-forming systems, it can be expected that taking into account the many-body nature of the MRO will help to elucidate also phenomena like the crystallization of liquids~\cite{kelton2010crystallization,rodrigues2023recent}, the structural origin of which is still far from being well understood. Notably, there is evidence that the local structure of the crystallization nucleus is sometimes different from the one of the crystalline phase, since the process is affected by entropic, packing, and elastic effects~\cite{tanaka2022order}. Thus, using an observable like the four-point correlation function that is able to probe the many-body nature of the MRO of the liquid approaching crystallization will allow to gain important insight into the microscopic mechanisms that trigger the crystallization process.

Overall, our findings indicate that the MRO in network glass-formers is rich and exhibit non-trivial temperature and composition dependence which is hardly detectable from standard (two-point) structural probes. Higher-order correlation functions such as the ones used in this work are expected to facilitate this discovery of hidden order within disorder, which is key for the establishment of a firm structure-relation in disordered materials.

\section{Acknowledgment}
We thank L.F. Ding for discussion. W.K. is a senior member of the Institut Universitaire de France.

\normalem  

%

\onecolumngrid

\end{document}